\documentclass[prb,aps,twocolumn,aip,jcp]{revtex4}
\usepackage{graphicx}
\usepackage{color}
\usepackage{calligra}
\usepackage{amsmath}
\usepackage{multirow}
\usepackage{float}

\begin{document}

\title{The correlation discrete variable representation revisited}
\author{Uwe Manthe$^*$}
\affiliation{Theoretische Chemie, Fakult\"at f\"ur Chemie, Universit\"at Bielefeld,
  Universit\"atsstr. 25, D-33615 Bielefeld, Germany}
\date{\today}

\begin{abstract}
The correlation discrete variable representation (CDVR) enables efficient
quantum dynamics calculation with the multi-layer multi-configurational
time-dependent Hartree (MCTDH) approach on general potential energy surfaces.
It employs a time-dependent quadrature to compute potential
energy matrix elements, thereby eliminating the need to refit the potential
to a sum of products form. The non-hierarchical CDVR conserves the inherent
symmetry properties of tree-shaped wavefunction representations and
drastically reduces the number of grid points compared to the original
hierarchical CDVR. However, it requires projection on the space
spanned by the single-hole functions (SHFs) at each node of the tree, which
can introduce unphysical couplings for unconverged basis sets.
In this work, the non-hierarchical CDVR is revisited and a revised
approach that avoids explicit projection on the single-hole space is introduced.
The computational costs of the revised approach scale favorably with the number
of single-particle functions (SPFs): for a tree with three edges at each node
and $n$ SPFs at each edge, a $n^4$ scaling is achieved. Furthermore,
a revised scheme that uses artificial SPFs to systematically increase the
accuracy of the CDVR quadrature is presented. Computations studying the
photodissociation of NOCl, the vibrational states of methyl, and the
non-adiabatic quantum dynamics of photoexcited pyrazine demonstrate
the accuracy and efficiency of the revised non-hierarchical CDVR.
Notably, for the 24-dimensional pyrazine system the use of the CDVR does not
increase the required CPU time compared to calculations utilizing
the sum of products form of the vibronic coupling model.
\\[5ex]

$^*$ Author to whom correspondence should be addressed:\\
uwe.manthe@uni-bielefeld.de
\end{abstract}

\maketitle
\newpage

\section{Introduction \label{sec1}}

The (multi-layer) multi-configurational time-dependent Hartree (MCTDH) approach
\cite{MMC,MMC1,WT3,M6} utilizes tensor contraction to enable accurate
high-dimensional quantum dynamics calculation. It has been used by many researchers
to study a large variety of systems.
\cite{MCTDHex2,MCTDHex4,MCTDHex5,MCTDHex7,WBRSM,MCTDHex8,MCTDHex9,MCTDHex10,MCTDHex13,MCTDHex14,MCTDHex15,MCTDHex17,MCTDHex18,MCTDHex19,MCTDHex20,MCTDHex23,MCTDHex24,MCTDHex25,MCTDHex26,NHoM,MCTDHex27,MCTDHex28,MCTDHex29,MCTDHex30,MCTDHex31}
High-dimensional MCTDH calculations on detailed full-dimensional potential energy
surfaces (PES) studied, e.g., reactions of methane with atoms, 
\cite{HM1,HM2,WWM,vHNM,NvHM,WeM5,WeM6,WeM8,ElM,WeM7,ElM2,ElM3,ElM4,ZM2,HoM},
proton transfer in malonaldehyde,  
\cite{CVM,HCVM,HaM1,MAMCTDH,HaM2,MAMCTDH2},
or large amplitude motion in the Zundel and Eigen cations.
\cite{H5O2+MCTDH,H5O2+MCTDH2,H5O2+MCTDH3,H5O2+MCTDH4,H5O2+MCTDH5,H5O2+MCTDH6,H9O4+MCTDH}
Benchmark MCTDH simulations based on high-dimensional model Hamiltonians
investigated the non-adiabatic dynamics in pyrazine \cite{WMC,WMC2,RWMC}
and the electron transfer and transport in the condensed phase \cite{WT3,WST,KCBWT,WPHT}.
Further examples can be found in various reviews on the subject (see, e.g., Refs.
\onlinecite{MCTDHreview,MCTDHreview2,HMreview1,MCTDHreview4,MCTDHbook,MCTDHreview3,Wreview,M9}).

The representation of the Hamiltonian is an important aspect in any quantum
dynamics method using tensor contraction. An efficient implementation of the
MCTDH approach can be straightforwardly achieved for Hamiltonians that can be
written as a sum of products (SOP) of operators acting only on a single degree
of freedom. The SOP form was intensively used in early work utilizing the
(single-layer) MCTDH approach.\cite{MMC1,MCTDHreview} In multi-layer MCTDH
calculations, matrix elements of SOP Hamiltonians can be  
recursively calculated using successive bottom up and to down sweeps.\cite{M6}

Studying the quantum dynamics on accurate ab initio PESs posts another challenge.
These PESs typically do not show a SOP form. Different strategies have been devised to
address the issue. Meyer and co-workers \cite{JM2,JM5,MGPotfit,MLPotfit,MCPotfit,CPPotfit}
developed techniques to refit the PES in the SOP form required by the standard 
MCTDH implementation. They successfully used this approach to study
malonaldehyde \cite{MAMCTDH,MAMCTDH2} and the Zundel and Eigen cations
\cite{H5O2+MCTDH,H5O2+MCTDH2,H5O2+MCTDH3,H5O2+MCTDH4,H5O2+MCTDH5,H5O2+MCTDH6,H9O4+MCTDH}.
Alternatively, grid-based based schemes can be used to directly compute potential energy
matrix elements without refitting the PES. The correlation discrete variable
representation (CVDR) \cite{M3,vHM2,vHM3,M6,M7} employs a time-dependent
quadrature to efficiently compute all potential energy matrix elements appearing
in the MCTDH equations of motion. This approach was successfully used, e.g.,
to study the reaction of methane with different atoms 
\cite{HM1,HM2,HM4,HM5,WWM,SM,vHNM,NvHM,WeM5,WeM6,WeM8,ElM,WeM7,ElM2,ElM3,ElM4,ZM2,WestPNM,SchM,SchM2}
or to compute the tunneling splittings in the ground and vibrationally excited states 
of malonaldehyde. \cite{CVM,HCVM,HaM1,HaM2} The collocation scheme of Wodraszka and
Carrington \cite{MCTDHcolloc1,MCTDHcolloc2,MCTDHcolloc3,MCTDHcolloc4}
offers an alternative grid-based approach facilitating MCTDH calculations on general potential
energy surfaces.

The original CDVR approach \cite{M3} designed for single MCTDH calculations can be
viewed as a two-layer Smolyak \cite{Smolyak1,Smolyak2,Smolyak3} quadrature. 
At the top layer, a sparse grid accounting for correlation effects is built based on
time-dependent discrete variable representations (DVRs) associated with the 
single-particle functions (SPFs). On the second layer, finer grids increase
the grid's density in a single direction using the respective time-independent
DVR points. This design scheme can be extended to a multi-layer
MCTDH/CDVR approach.\cite{M6} It employs increasingly large grids at
lower layers of the multi-layered tree representation and thus
is called hierarchical CDVR in the following.

Progress in recent MCTDH method development took advantage of a symmetric,
non-hierarchical view on the multi-layer MCTDH wavefunction. By altering the
choice of the root node, the role of SPFs and single-hole functions (SHF)
in the MCTDH formalism can be interchanged and different equivalent
multi-layer MCTDH representation can be obtained.\cite{LubichInt,MCTDHLarsson,WeiM2}
Based on these ideas, a non-hierarchical CDVR approach has been 
introduced.\cite{ElM5,ElHoM2} The approach employs a revised grid
design that uses roughly equal numbers of grid point at all layers of the
tree. Total number of grid points is significantly decreased compared
to the original hierarchical CDVR and roughly equals the total number of
expansion coefficients used in the MCTDH wavefunction representation.

The non-hierarchical CDVR introduced in Refs. \onlinecite{ElM5,ElHoM2}
requires projection onto the space spanned by the SHF to obtain edge-based
potential corrections. This projection introduces undesirable features.\cite{ElHoM2}
First, the CDVR potential operator can affect even parts of the wavefunction
that depend on coordinates not present in potential function. This unphysical
behavior only vanishes in the limit of a converged MCTDH wavefunction representation.
Second, the CDVR scheme does not guarantee a correct description of separable
potentials in case of an unconverged MCTDH wavefunction representation.
Third, the calculation of the CDVR matrix elements scales less favorably than
the application of the CDVR potential operator: for a node with $f$ edges
and $n$ SPFs at each edge, the computational costs for the calculation of the
CDVR matrix elements scale as $n^{f+3}$ while the application of the
CDVR potential operator exhibits a more favorable $n^{f+1}$ scaling.

In the present work, a revised non-hierarchical CDVR is developed that
avoids the projection onto the space spanned by the SHF and eliminates all
the undesirable features mentioned above. The result scheme can implemented
with a numerical effort scaling as $n^{f+1}$.

Furthermore, a scheme to replace weakly occupied SPFs by artifical SPFs
designed to increase the accuracy of CDVR quadrature is presented. This scheme
is based upon the ideas outlined in Ref. \onlinecite{ElHoM} and enables
numerically exact MCTDH/CDVR calculations that properly converge to the
exact result with increasing SPF basis set size.

The article is organized as follows. The multi-layer MCTDH approach and
the original non-hierarchical CDVR are briefly in Sects. \ref{sec2}
and \ref{sec3}. Then the revised CDVR scheme is
decribed in Sect. \ref{sec4}. Sect.\ref{sec5} discusses the use
of artifical SPFs to improve the accuracy of the CDVR quadrature.
Numerical examples illustrate the accuracy and efficency of the different
schemes (Sect.\ref{sec6}): the photodissociation of NOCl (3D),
the vibrational states of methyl (6D), and the S$_0$$\to$S$_2$ excitation
in pyrazine (24D) are studied. Finally, conclusions and an outlook are given
in Sect.\ref{sec7}.

\section{The multi-layer MCTDH approach \label{sec2}}

\begin{figure}[t]
\caption{The multi-layer MCTDH approach: the grouping of physical coordinates
$x_i$ into logical coordinates ${\bf q}^{l;\kappa_1,..,\kappa_{l-1}}$ and
the labeling of notes and edges are illustrated.}
\includegraphics[width=0.45\textwidth]{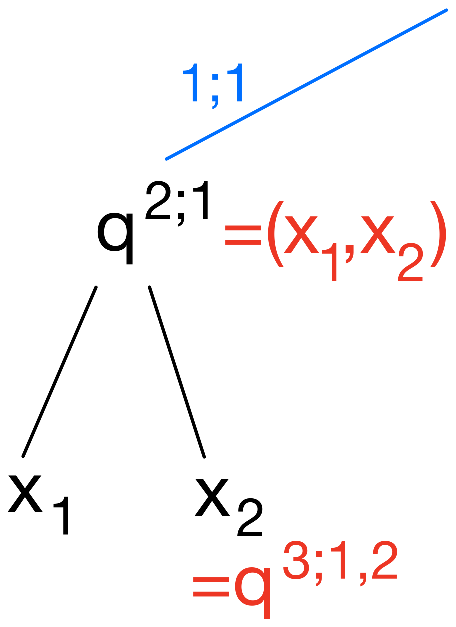}
\end{figure}

The multi-layer MCTDH approach \cite{MMC,MMC1,WT3,M6} employs a tree-shaped
structure to represent the system's wavefunction. As illustrated in Fig.1,
the physical coordinates $x_1,x_2,..,x_f$ are grouped into sets of
logical coordinates ${\bf q}^{l;\kappa_1,..,\kappa_{l-1}}$ where the
superscripts $(l;\kappa_1,..,\kappa_{l-1})$ indicate the specific node and
$l$ is the number of the layer.
An edge connecting the nodes $(l;\kappa_1,..,\kappa_{l-1})$
and $(l+1;\kappa_1,..,\kappa_{l})$ is denoted by 
$(l;\kappa_1,..,\kappa_{l})$ (see the colored labels in Fig.1).

Starting from the root node of the tree, the system's wavefunction
$\Psi(x_1,..,x_f,t)$ is expanded in orthonormal sets of SPFs
that can be associated with the $\nu$ edges $(1;\kappa)$ connecting
the root node with its children:
\begin{equation}
\Psi(x_1,..,x_f,t) = 
\sum_{j_1,..,j_\nu} A^{1}_{j_1,..,j_\nu}(t) 
\prod_{\kappa=1}^\nu \phi_{j_\kappa}^{1;\kappa}({\bf q}^{2,\kappa},t) ~.
\label{eq1}   
\end{equation}
The SPFs are defined recursively. The orthonormal set of SPFs
$\phi_{n}^{l;\kappa_1,..,\kappa_l}({\bf q}^{l+1;\kappa_1,..,\kappa_{l}},t)$
at the edge $(l;\kappa_1,..,\kappa_l)$ is obtained from the SPFs
at the lower-layer edges  $(l+1;\kappa_1,..,\kappa_l,\kappa_{l+1})$:
\begin{equation}
\phi_{n}^{l;\kappa_1,..,\kappa_l}
= \sum_{j_1,..,j_\nu} 
A^{l+1;\kappa_1,..,\kappa_l}_{n;j_1,..j_\nu} 
\prod_{\kappa=1}^\nu
\phi_{j_\kappa}^{l+1;\kappa_1,..,\kappa_{l},\kappa}~.
\label{eq2}
\end{equation}
Here the $A^{l+1;\kappa_1,..,\kappa_l}$ denotes expansion coefficients
and $\nu$ is the number of edges connecting the node $(l;\kappa_1,..,\kappa_{l-1})$
to nodes $(l+1;\kappa_1,..,\kappa_{l})$ of the lower layer.
At the bottom layer of the tree, the SPFs are replaced
by time-independent basis functions $\chi_{n}^{\kappa}$. 

In the above description, all nodes and edges are labeled according to their
absolute position in the given tree. These labels specify the positions relative
to the chosen root node. However, the definitions of all objects
appearing in the above description only refer to neighboring nodes and edges.
Thus, the relative position in the tree play a central role.
To address notes and edges via their relative position, an alternative notation
specifying the neighborhood of an arbitarily chosen reference node
$(\lambda)=(l;\kappa_1,..,\kappa_{l-1})$ is used.\cite{ElHoM2}
Nodes connected to $(\lambda)$ are denoted as $(\lambda \circ j)$:
$(\lambda \circ 0)$ refers to the node $(l-1;\kappa_1,..,\kappa_{l-2})$
directly above $\lambda$ while $(\lambda \circ \kappa)$ with $\kappa>0$ 
refers to a lower node $(l+1;\kappa_1,..,\kappa_{l-1},\kappa)$. An edge
connecting the nodes $(\lambda)$ and $(\lambda \circ \kappa)$ can be
denoted either by $(\lambda | \lambda \circ \kappa)$ or by
$(\lambda \circ \kappa| \lambda)$. Here the ordering allows one to
distinguish directions: $(\lambda | \nu )$ implies
moving from node $(\lambda)$ to node $(\nu)$ while $(\nu | \lambda)$
specifies the opposite direction. Using this notation, Eq.(\ref{eq2}) can be
rewritten as
\begin{equation}
\phi_{n}^{\lambda \circ 0 | \lambda} = \sum_{j_1,..,j_\nu} 
A^{\lambda}_{n;j_1,..j_\nu} 
\prod_{\kappa=1}^\nu
\phi_{j_\kappa}^{\lambda | \lambda \circ \kappa}~.
\label{eq3}
\end{equation}
where a downward direction is used in the superscripts of the SPFs.

At any edge $(\lambda | \lambda \circ \kappa)$, the wavefunction
can be decomposed in a sum of SPFs $\phi_{n}^{\lambda | \lambda \circ \kappa}$ and 
single-hole functions (SHFs) $\Psi_{n}^{\lambda | \lambda \circ \kappa}$:
\begin{equation}
\Psi = \sum_n \phi_{n}^{\lambda | \lambda \circ \kappa} \cdot
\Psi_{n}^{\lambda | \lambda \circ \kappa} ~.
\label{eq4}   
\end{equation}
The SHFs $\Psi_{n}^{\lambda | \lambda \circ \kappa}$ depend on all coordinates
not included in SPFs $\phi_n^{\lambda | \lambda \circ \kappa}$.
Importantly, the SPFs and SHFs obey different normalization conditions:
\begin{align}
\left< {\phi}_{n}^{\lambda | \lambda \circ \kappa} \middle|
{\phi}_{m}^{\lambda | \lambda \circ \kappa} \right> &= \delta_{mn} ~,\nonumber\\
\left< {\Psi}_{n}^{\lambda | \lambda \circ \kappa} \middle|
{\Psi}_{m}^{\lambda | \lambda \circ \kappa} \right> &= \rho_{nm}^{\lambda | \lambda \circ \kappa}~,
\label{eq5}   
\end{align}
where ${\boldsymbol \rho}^{\lambda | \lambda \circ \kappa}$ is the single-particle
density matrix associated with the edge $(\lambda | \lambda \circ \kappa)$.

In the above description, a specific node was selected as root node and
associated with the system's wavefunction (see Eq.(\ref{eq1})). However,
only the tree's connectivity is important and the physical wavefunction
is invariant with respect to the choice of the root node.\cite{MCTDHLarsson}
Altering the root node potentially exchanges the roles of SPFs and
SHFs Within the present notation, any change in the directionality of an edge
affects the superscripts used in the SPFs and SHFs. Transformed SHFs
and transformed SPFs as defined in Ref.\onlinecite{WeiM2} can thus be denoted
as SPFs and SHFs where the directionality in the superscript is inverted:\cite{ElHoM2}
\begin{align}
\phi_{n}^{\lambda \circ \kappa| \lambda}&=
\sum_k (\rho^{\lambda |\lambda \circ \kappa})^{-\frac{1}{2}}_{kn}
\Psi_{k}^{\lambda |\lambda \circ \kappa} ~,
\nonumber\\
\Psi_{n}^{\lambda \circ \kappa| \lambda}&=
\sum_k (\rho^{\lambda |\lambda \circ \kappa})^{\frac{1}{2}}_{nk}
\phi^{\lambda |\lambda \circ \kappa}_{k} ~.
\label{eq6}
\end{align}

Using this notation, the edge-based representation of the wavefunction takes the
form 
\begin{align}
\Psi &= \sum_{n,m} \left(\rho^{\lambda |\lambda \circ \kappa}\right)^{\frac{1}{2}}_{nm} \cdot
\phi_{m}^{\lambda \circ \kappa | \lambda} 
\cdot \phi_n^{\lambda |\lambda \circ \kappa} ~,
\nonumber\\
&= \sum_{n,m}  \left(\rho^{\lambda \circ \kappa | \lambda}\right)^{\frac{1}{2}}_{nm} \cdot
\phi_{m}^{\lambda | \lambda \circ \kappa}
\cdot \phi_n^{\lambda \circ \kappa| \lambda} ~.
\label{eq7}   
\end{align}
Consequently,
\begin{equation}
\rho^{\lambda|\lambda \circ \kappa}_{nm}=\rho^{\lambda \circ \kappa|\lambda}_{mn}~.
\label{eq8}   
\end{equation}
Defining the configurations $\Phi^\lambda_J$ as
\begin{equation}
\Phi^\lambda_J = \prod_\kappa \phi_{j_\kappa}^{\lambda |\lambda \circ \kappa} ~
\label{eq9}   
\end{equation}
with $\kappa=0,..,\nu$ for non-root nodes and $\kappa=1,..,\nu$ for the root node,
the wavefunction can be expanded in the basis of the configurations at any non-root node
as
\begin{align}
\Psi &= \sum_{J} \tilde{A}^{\lambda}_{J} \cdot \Phi^\lambda_J ~, \nonumber\\
\tilde{A}^{\lambda}_{J} &= \sum_{n} (\rho^{\lambda \circ 0 | \lambda})^{\frac{1}{2}}_{j_0 n}
\cdot A^{\lambda}_{n;j_1,..j_\nu}
\label{eq10}   
\end{align}
where the multi-index $J=(j_0,j_1,..j_\nu)$ is used.

The MCTDH equations of motion for the coefficients
$A^\lambda_J=A^\lambda_{j_0;j_1,j_2,..}$ 
then read \cite{ElHoM2} ($\hat{H}$ is the system's Hamiltonian)
\begin{equation}
i\frac{\partial}{\partial t} A^\lambda_J =
\langle \Phi^\lambda_J| \hat{H} | \Psi \rangle
\label{eq11}   
\end{equation}
if $\lambda$ is the root node (with $j_0=1$) and
\begin{eqnarray}
i\frac{\partial}{\partial t} A^\lambda_J &=&
\left(1-\sum_{m} |\phi^{\lambda \circ 0|\lambda}_m \rangle
\langle \phi^{\lambda \circ 0|\lambda}_m | \right)
\nonumber\\ &&\cdot
\sum_n (\rho^{\lambda \circ 0 |\lambda})^{-\frac{1}{2}}_{j_0 n} \cdot
\langle\Phi^\lambda_{n,j_1,j_2,..}| \hat{H} | \Psi \rangle
\label{eq12}   
\end{eqnarray}
if $\lambda$ is a non-root node.

The matrix elements $\langle \Phi^\lambda_J| \hat{H} | \Psi \rangle$ 
appearing in the equations of motion can be straightforwardly and 
efficiently computed for Hamiltonians $\hat{H}_{SOP}$ showing a 
sum of products form\cite{MMC,M6}
\begin{equation}
\hat{H}_{SOP} = 
\sum_{t=1}^s c_t \cdot \prod_{k=1}^{f} \hat{h}^{prim;k}_t 
\label{eq13}
\end{equation}
where $\hat{h}^{prim;k}_t~$ denotes an operator acting only on the 
(primitive) coordinate $x_k$. Then at each node $(\lambda)$
the terms $\langle \Phi^\lambda_J| \hat{H} | \Psi \rangle$
appearing in the equations of motion can be efficiently computed
via
\begin{align}
\langle \Phi^\lambda_J| \hat{H} | \Psi \rangle =
\sum_{t=1}^s c_t \cdot
& \sum_{i_0}          
\langle \phi^{\lambda |\lambda \circ 0}_{j_0} |\hat{h}^{\lambda;0}_t 
| \phi^{\lambda |\lambda \circ 0}_{i_0} \rangle 
\nonumber\\ & \cdot \sum_{i_1}          
\langle \phi^{\lambda |\lambda \circ 1}_{j_1} |\hat{h}^{\lambda;1}_t 
| \phi^{\lambda |\lambda \circ 1}_{i_1} \rangle 
\nonumber\\ &  \cdot ... 
\nonumber\\ & \cdot \sum_{i_\nu}          
\langle \phi^{\lambda |\lambda \circ \nu}_{j_\nu} |\hat{h}^{\lambda;\nu}_t 
| \phi^{\lambda |\lambda \circ \nu}_{i_\nu} \rangle 
\cdot \tilde{A}^\lambda_{i_0,i_1,..,i_\nu} ~.
\label{eq14}
\end{align}                                              
Here the operator $\hat{h}^{\lambda;\kappa}_t$ is defined as 
the product of all operators $\hat{h}^{prim;l}_t$ with 
$x_l \in q^{\lambda \circ \kappa}$ for $\kappa \ne 0$ or 
$x_l \notin q^{\lambda}$ for $\kappa=0$. All matrix elements 
$\langle \phi^{\lambda |\lambda \circ \kappa}_{j_\kappa} |\hat{h}^{\lambda;\kappa}_t 
| \phi^{\lambda |\lambda \circ \kappa}_{i_\kappa} \rangle$
can be efficiently calculated using the recursive scheme described in
Ref.\onlinecite{M6}. 

\section{Non-hierarchical multi-layer CDVR \label{sec3}}

The CDVR approach extends the type of Hamiltonians which can be efficiently
studied by MCTDH calculations. It enables the description of more general
Hamiltonians of the form
\begin{align}
\hat{H} &= \sum_{t=1}^s V_t(x_1,x_2,..,x_{l_t}) \cdot \hat{H}_t ~,
\nonumber\\
\hat{H}_t &= \prod_{d={l_t+1}}^{f} \hat{h}^{prim;d}_t
\label{eq15}
\end{align}
which simultaneously include multi-dimensional functions
\mbox{$V_t(x_1,x_2,..,x_l)$} and products of 
single-particle operators in the same term.\cite{M3,VENDM,M6}
The matrix elements of these functions are computed using time-dependent 
quadratures based on grids obtained by diagonalizing the matrix
representations of coordinate operators in the SPF bases.

Grid representations associated with the
$\phi^{\lambda | \lambda \circ \kappa}_j(x_a, x_b, ..)$
are obtained by (approximate simultaneous) diagonalization of the
coordinate matrices \cite{M3,vHM3}
\begin{equation}
\langle \phi^{\lambda | \lambda \circ \kappa}_{j} | \hat{x}_n |
\phi^{\lambda | \lambda \circ \kappa}_{i}\rangle ~,~ 
n=a, b, .. ~. 
\label{eq16} 
\end{equation}
The unitary matrix transforming to the (approximately)
diagonal representation defines the basis-grid transformation 
$\langle \xi^{\lambda | \lambda \circ \kappa}_m | 
\phi^{\lambda | \lambda \circ \kappa}_i \rangle$ and
the grid points ${\boldsymbol \xi}^{\lambda | \lambda \circ \kappa}_m$ 
are given by
\begin{equation}
(\xi_n)^{\lambda | \lambda \circ \kappa}_m=
\langle \xi^{\lambda | \lambda \circ \kappa}_{m} | \hat{x}_n |
\xi^{\lambda | \lambda \circ \kappa}_{m}\rangle ~,~  n=a, b, .. ~. 
\label{eq17}
\end{equation}
Localized SPFs $\xi^{\lambda | \lambda \circ \kappa}_m(x_a, x_b, .., t)$
associated with the grid points
${\boldsymbol \xi}^{\lambda | \lambda \circ \kappa}_m$
are defined by
\begin{align}
|\xi^{\lambda | \lambda \circ \kappa}_m \rangle = 
\sum_j |\phi^{\lambda | \lambda \circ \kappa}_{j} \rangle \cdot
\langle\phi^{\lambda | \lambda \circ \kappa}_{j} | 
\xi^{\lambda | \lambda \circ \kappa}_m \rangle ~.
\label{eq18}
\end{align}
Since the different coordinate matrices generally do not commute, only
approximate simultaneous diagonalization is feasible for multi-dimensional
$\phi^{\lambda | \lambda \circ \kappa}_j$ and appropriate weighting and 
scaling of the coordinate matrices is important.\cite{M7,ElM5}

The non-hierarchical CDVR approach employs operators $\hat{V}^{CDVR}_t$
that include contributions from the grid representations associated
with all nodes and edges to simulate the action of the potential
$V_t(x_1,x_2,..,x_{l_t})$ on a wavefunction.\cite{ElM5,ElHoM}
The contribution from a node $(\lambda)$ is given by
\begin{equation}
\hat{V}_t^\lambda = \sum_{N} | \Xi^\lambda_N \rangle
V_t({\boldsymbol \Xi}^\lambda_N)
\langle \Xi^\lambda_N|
\label{eq19}
\end{equation}
where                                                            
\begin{equation} 
| \Xi^\lambda_N \rangle = | \prod_{\kappa=0}^\nu 
\xi_{j_\kappa}^{\lambda |\lambda \circ \kappa} \rangle 
\label{eq20}
\end{equation}
are grid-transformed configurations and 
\begin{equation}
{\boldsymbol \Xi}^\lambda_N=({\boldsymbol \xi}_{n_0}^{\lambda |\lambda \circ 0},
{\boldsymbol \xi}_{n_1}^{\lambda |\lambda \circ 1}, 
{\boldsymbol \xi}_{n_2}^{\lambda |\lambda \circ 1},..)
\label{eq21}
\end{equation}
defines a complete set of coordinate values. The contribution from
an edge $(\lambda \circ 0|\lambda)$,
\begin{equation}
\hat{V}_t^{\lambda \circ 0|\lambda} = 
\hat{V}_t^{\lambda |\lambda \circ 0} =
\sum_{n,m} | \Xi^{\lambda \circ 0|\lambda}_{nm} \rangle
V_t({\boldsymbol \Xi}^{\lambda \circ 0|\lambda}_{n,m})
\langle \Xi^{\lambda \circ 0|\lambda}_{nm}| ~,
\label{eq22}
\end{equation}
is defined using grid-transformed edge-based configurations
\begin{equation} 
| \Xi^{\lambda \circ 0|\lambda}_{nm} \rangle = 
| \Xi^{\lambda | \lambda \circ 0}_{mn} \rangle = 
| \xi^{\lambda \circ 0 |\lambda}_n \cdot
\xi^{\lambda |\lambda \circ 0}_m \rangle 
\label{eq23}
\end{equation}
and the associated set of coordinates
\begin{equation}
{\boldsymbol \Xi}^{\lambda \circ 0|\lambda}_{n,m}=
{\boldsymbol \Xi}^{\lambda |  \lambda \circ 0}_{m,n}=
({\boldsymbol \xi}_{n}^{\lambda \circ 0|\lambda},
{\boldsymbol \xi}_{m}^{\lambda |\lambda \circ 0}) ~.
\label{eq24}
\end{equation}
The CDVR potential operator is given by
\begin{equation}
\hat{V}^{CDVR}_t = \sum_{\lambda \in \Lambda} \hat{V}_t^\lambda - 
\sum_{\substack{\lambda \in \Lambda \\ \lambda \neq root}} 
\hat{V}_t^{\lambda \circ 0 |\lambda}  
\label{eq25}
\end{equation}
where $\Lambda$ denotes the set of all nodes.

To obtain the matrix elements
$\langle \Xi^\lambda_I | V_t | \Psi \rangle$,
the sum in Eq.(\ref{eq25}) is rewritten separating contributions
of the node $\lambda$ and the $\nu+1$ subtrees originating from the 
$\nu+1$ edges connected to $\lambda$:   
\begin{equation}
\hat{V}_t^{CDVR} = \hat{V}_t^\lambda +
\sum_{\kappa=0}^\nu \Delta \hat{V}_t^{\lambda|\lambda \circ \kappa}~.
\label{eq26}
\end{equation}
The resulting expression reads \cite{ElHoM2}
\begin{align}
\langle \Xi^\lambda_I | V_t | \Psi \rangle 
=& V_t({\boldsymbol \Xi}^\lambda_I) \cdot \langle \Xi^\lambda_I | \Psi \rangle
\nonumber \\
& + \sum_{\kappa} \sum_{l}
\langle \Xi^\lambda_I |\Xi^{\lambda |\lambda \circ \kappa}_{i_\kappa l} \rangle
\nonumber \\
& ~~ \cdot \sum_{j_\kappa,k} 
\langle \Xi^{\lambda |\lambda \circ \kappa}_{i_\kappa l}|\Delta
\hat{V}_t^{\lambda|\lambda \circ \kappa}
| \Xi^{\lambda |\lambda \circ \kappa}_{j_\kappa k} \rangle 
\nonumber \\
& ~~ \cdot \sum_{J_{/\kappa}} 
\langle \Xi^{\lambda |\lambda \circ \kappa}_{j_\kappa k} | \Xi^\lambda_J \rangle
\cdot \langle \Xi^\lambda_J | \Psi \rangle ~.
\label{eq27}
\end{align}
where $J_{/\kappa}$ is short hand notation for $(j_0,..,j_{\kappa-1},j_{\kappa+1}, ..)$. 
All calculations required in the above equation can be implemented with a
numerical effort scaling not worse than either $n \cdot N$ or $n^4$ where $n$ 
denotes the number of SPFs associated with an edge and $N$ the number of configurations
associated with a node. Since $N=n^d$ for a node with $d$ edges and
non-redundant nodes show at least three nodes, a $n^4$ scaling is obtained for
an optimally designed tree.

In Eq.(\ref{eq27}), the matrix elements representing the edge-based contribution
$\Delta \hat{V}_t^{\lambda|\lambda \circ \kappa}$ to the potential operator
in the basis of the edge-basis configuations $\Xi^{\lambda |\lambda \circ \kappa}_{nm}$, 
\begin{align}
\Delta V_{t;iljk}^{\lambda|\lambda \circ \kappa} =
\langle \Xi^{\lambda |\lambda \circ \kappa}_{il}|\Delta
\hat{V}_t^{\lambda|\lambda \circ \kappa}
| \Xi^{\lambda |\lambda \circ \kappa}_{jk} \rangle ~,
\label{eq28}
\end{align}
take a central role. They describes the CDVR correction to the 
potential operator $\hat{V}_t^\lambda$ which only considers the grid
${\boldsymbol \Xi}^\lambda_N$ associated with the specific node. The recursive
calculation of these matrix elements is particularly demanding. Introducing a modified
potential operator 
\begin{equation}
\hat{V}^\lambda_{t,\kappa} = \hat{V}_t^\lambda +
\sum_{\substack{\mu \\ \mu \neq \kappa}} \Delta \hat{V}_t^{\lambda|\lambda \circ \mu}
\label{eq29}
\end{equation}
where a single correction term $\hat{V}^{\lambda|\lambda \circ \kappa}$ is omitted
in $\hat{V}_t$, these matrix element can
be computed by \cite{ElHoM2}
\begin{align}
\Delta V_{t;nlmk}^{\lambda \circ \kappa|\lambda} =&
- V_t({\boldsymbol \Xi}^{\lambda \circ \kappa|\lambda}_{n,l}) 
\cdot \delta_{nm} \cdot \delta_{lk}
\nonumber\\
&+ \sum_I \langle \Xi^{\lambda \circ \kappa|\lambda}_{nl}|\Xi^{\lambda}_I \rangle
\cdot \langle \Xi^{\lambda}_I | \hat{V}^{\lambda}_{t,\kappa}|
\Xi^{\lambda \circ \kappa|\lambda}_{mk} \rangle ~.
\label{eq30}
\end{align}
Here for any index pair $m,k$ the matrix element
$\langle \Xi^\lambda_I | \hat{V}^\lambda_{t,\kappa}|
\Xi^{\lambda \circ \kappa|\lambda}_{m,k} \rangle$ can be calculated analogously
to the matrix element $\langle \Xi^\lambda_I | V_t | \Psi \rangle$ on the left
hand side of Eq.(\ref{eq27}). Thus, the numerical effort required for the
calculation of the $\Delta \hat{V}_{t;lknm}^{\lambda \circ 0|\lambda}$
scales as $n^2$ times (where $n$ denotes the number of SPFs associated with an edge)
the numerical effort required for the evalution of Eq.(\ref{eq27}), resulting
a $n^6$ scaling for an optimally designed tree.

This $n^6$ scaling can be viewed as one important disadvantage of the CDVR
approach: conventional MCTDH calculations relying on a SOP form of the
Hamiltionian achieve a $n^4$ scaling for optimally designed trees. Another
disadvantage of the non-hierarchical CDVR scheme described above is the appearance
of the node-edge transformation, described by the matrices
$\langle \Xi^{\lambda |\lambda \circ \kappa}_{j_\kappa k} | \Xi^\lambda_J \rangle$,
in Eq.(\ref{eq27}). This transformation affects all coordinates of the system.
Consequently, the CDVR-potential operator effectively acts not only
on the $l$ coordinates present in \mbox{$V_t(x_1,x_2,..,x_l)$} but on all coordinates
of the system. This undesirable effect and different schemes to mitigate the
consequences have been discussed in detail in Ref.\onlinecite{ElHoM2}. Both
shortcomings will addressed by the revised CDVR scheme described in the following.

\section{A revised non-hierarchical CDVR \label{sec4}}

The issues discussed above can be addressed by replacing the
edge-based contribution to the potential operator appearing in Eq.(\ref{eq27}),
\begin{align}
\Delta \hat{V}_{t}^{\lambda|\lambda \circ \kappa}= \sum_{i,l,j,k}
| \Xi^{\lambda |\lambda \circ \kappa}_{jk} \rangle \cdot
\Delta V_{t;iljk}^{\lambda|\lambda \circ \kappa} \cdot
\langle \Xi^{\lambda |\lambda \circ \kappa}_{il}| ~,
\label{eq31}
\end{align}
by the revised operator 
\begin{align}
\Delta \hat{V}_{t,rev}^{\lambda|\lambda \circ \kappa}= \sum_{i,j}
| \xi^{\lambda |\lambda \circ \kappa}_{i} \rangle \cdot
\Delta V_{t;ij}^{\lambda|\lambda \circ \kappa} \cdot
\langle \xi^{\lambda |\lambda \circ \kappa}_{j}| ~.
\label{eq32}
\end{align}
The revised operator no longer refers to the transformed SHFs
$\xi^{\lambda \circ \kappa | \lambda}_{i}$ and thus
avoids projection onto the space spanned by the SHF. Consequently,
the revised operator of Eq.(\ref{eq32}) acts for $\kappa>0$ only on
the $l$ coordinates present in \mbox{$V_t(x_1,x_2,..,x_l)$} and does
not affect any other coordinates.

The revised operator is chosen to optimally approximate the 
original operator $\Delta \hat{V}_{t}^{\lambda|\lambda \circ \kappa}$
in
\begin{align}
V_t(x_1,x_2,..,x_{l_t}) \cdot \hat{H}_t | \Psi \rangle ~.
\label{eq33}
\end{align}
It is obtained by minimizing
\begin{align}
\left\| \left( \Delta \hat{V}_{t}^{\lambda|\lambda \circ \kappa}
- \Delta \hat{V}_{t,rev}^{\lambda|\lambda \circ \kappa} \right)
\hat{H}_t \Psi \right\|^2 ~  
\label{eq34}
\end{align}
with respect to the $\Delta \hat{V}_{t;ij}^{\lambda|\lambda \circ \kappa}$.
As described in detail in Appendix A, the variation yields
\begin{align}
\Delta V_{t;ij}^{\lambda|\lambda \circ \kappa} = &
\sum_{n} u_{t;jn}^{\lambda|\lambda \circ \kappa} \sum_{m} 
{u_{t;im}^{\lambda|\lambda \circ \kappa}}^* \cdot
\frac{1}{r_{t;n}^{\lambda|\lambda \circ \kappa}+
         r_{t;m}^{\lambda|\lambda \circ \kappa}}
\nonumber\\
& \cdot \sum_{l} {u_{t;ln}^{\lambda|\lambda \circ \kappa}}^* 
\sum_{k} u_{t;km}^{\lambda|\lambda \circ \kappa} 
\left({X_{t;kl}^{\lambda|\lambda \circ \kappa}}^*+
X_{t;lk}^{\lambda|\lambda \circ \kappa} \right)
\label{eq35}
\end{align}
with
\begin{align}
X_{t;lk}^{\lambda|\lambda \circ \kappa}=&  
\sum_i \langle \hat{H}_t \Psi | \Xi_{li}^{\lambda|\lambda \circ \kappa} \rangle
\sum_{j,n} \Delta \hat{V}_{t;kinj}^{\lambda|\lambda \circ \kappa}
\langle \Xi_{nj}^{\lambda|\lambda \circ \kappa} | \hat{H}_t \Psi \rangle ~.
\label{eq36}
\end{align}
The $r_{t;n}^{\lambda|\lambda \circ \kappa}$ and
$u_{t;in}^{\lambda|\lambda \circ \kappa}$ in the Eq.(\ref{eq35})
are eigenvalues and eigenvectors defined by the eigenvalue problem
\begin{align}
\left\langle \hat{H}_t \Psi \middle| \xi_{i}^{\lambda|\lambda \circ \kappa} \rangle
\langle \xi_{j}^{\lambda|\lambda \circ \kappa} \middle| \hat{H}_t \Psi \right\rangle
= \sum_n u_{t;in}^{\lambda|\lambda \circ \kappa} \cdot r_{t;n}^{\lambda|\lambda \circ \kappa}
\cdot {u_{t;jn}^{\lambda|\lambda \circ \kappa}}^* .
\label{eq37}
\end{align}
 
Replacing $\Delta \hat{V}_{t}^{\lambda|\lambda \circ \kappa}$ by
$\Delta \hat{V}_{t,rev}^{\lambda|\lambda \circ \kappa}$ in Eq.(\ref{eq27}),
the action of a term $V_t \cdot \hat{H}_t$ in the Hamiltonian (\ref{eq15})
on the wavefunction at node $(\lambda)$ can be evaluated by
\begin{align}
\langle \Xi^\lambda_I | \hat{V}_t | \hat{H}_t \Psi \rangle 
=& V_t({\boldsymbol \Xi}^\lambda_I) \cdot \langle \Xi^\lambda_I |
\hat{H}_t \Psi \rangle \nonumber \\
& + \sum_{\kappa} \sum_{j_{\kappa}}
\Delta V_{t;i_\kappa j_{\kappa}}^{\lambda|\lambda \circ \kappa}
\cdot \langle \Xi^\lambda_{I_{/\kappa}, j_{\kappa}} | \hat{H}_t \Psi \rangle
\label{eq38}
\end{align}
where $(I_{/\kappa}, j_{\kappa})$ is short hand notation for
$(i_0,..,i_{\kappa-1},j_{\kappa},i_{\kappa+1}, ..)$.
In contrast to Eq.(\ref{eq27}), this equation no longer includes any
projection onto spaces spanned by the edge-based configurations
$\Xi^{\lambda |\lambda \circ \kappa}_{n m}$.

The CDVR corrections appear in Eq.(\ref{eq38}) in form of the
matrices ${\boldsymbol \Delta} \hat{\mathbf{V}}_{t}
^{\lambda|\lambda \circ \kappa}$. These can be straightforwardly
computed via Eq.(\ref{eq35}) once the matrices
$\mathbf{X_{t}}^{\lambda|\lambda \circ \kappa}$ defined in Eq.(\ref{eq36})
are known. The required $\mathbf{X_{t}}^{\lambda \circ \kappa|\lambda}$
(the change in the superscripts is required for notational reasons),
\begin{align}
X_{t;lk}^{\lambda \circ \kappa|\lambda}=&  
\sum_i \langle \hat{H}_t \Psi | \Xi_{li}^{\lambda \circ \kappa|\lambda} \rangle
\sum_{j,n} \Delta \hat{V}_{t;kinj}^{\lambda \circ \kappa|\lambda}
\langle \Xi_{nj}^{\lambda \circ \kappa|\lambda} | \hat{H}_t \Psi \rangle ~,
\label{eq39}
\end{align}
are obtained by inserting Eq.(\ref{eq30}) into Eq.(\ref{eq39}):
\begin{align}
X_{t;lk}^{\lambda \circ \kappa|\lambda}=&  
- \sum_i \langle \hat{H}_t \Psi | \Xi_{li}^{\lambda \circ \kappa|\lambda} \rangle
V_t({\boldsymbol \Xi}^{\lambda \circ \kappa | \lambda}_{k,i})
\langle \Xi_{ki}^{\lambda \circ \kappa|\lambda} | \hat{H}_t \Psi \rangle 
\nonumber\\ &
+ \sum_i \langle \hat{H}_t \Psi | \Xi_{li}^{\lambda \circ \kappa|\lambda} \rangle
\sum_{j,n}
\sum_I \langle \Xi^{\lambda \circ \kappa|\lambda}_{ki}|\Xi^{\lambda}_I \rangle
\nonumber\\ & \quad
\cdot \langle \Xi^{\lambda}_I | \hat{V}^{\lambda}_{t,\kappa}|
\Xi^{\lambda \circ \kappa|\lambda}_{nj} \rangle
\langle \Xi_{nj}^{\lambda \circ \kappa|\lambda} | \hat{H}_t \Psi \rangle
\nonumber\\ =&  
- \sum_i \langle \hat{H}_t \Psi | \Xi_{li}^{\lambda \circ \kappa|\lambda} \rangle
V_t({\boldsymbol \Xi}^{\lambda \circ \kappa|\lambda}_{k,i})
\langle \Xi_{ki}^{\lambda \circ \kappa|\lambda} | \hat{H}_t \Psi \rangle 
\nonumber\\ &
+ \sum_i \langle \hat{H}_t \Psi | \Xi_{li}^{\lambda \circ \kappa|\lambda} \rangle
\sum_{I_{/\kappa}}
\langle \Xi^{\lambda \circ \kappa|\lambda}_{ki}|\Xi^{\lambda}_{I_{/\kappa} i} \rangle
\nonumber\\ & \quad
\cdot \langle \Xi^{\lambda}_{I_{/\kappa} i} | \hat{V}^{\lambda}_{t,\kappa} 
\left( \sum_{n,j} |\Xi^{\lambda \circ \kappa|\lambda}_{nj} \rangle
\langle \Xi_{nj}^{\lambda \circ \kappa|\lambda} | \hat{H}_t \Psi \rangle \right)~.
\label{eq40}
\end{align}

In contrast to Eq.(\ref{eq30}), here $\hat{V}^{\lambda}_{t,\kappa}$ acts only
on a single wavefunction $\tilde{\Psi}$ given by 
\begin{align}
\tilde{\Psi} = \sum_{n,j} |\Xi^{\lambda \circ \kappa|\lambda}_{nj} \rangle
\langle \Xi_{nj}^{\lambda \circ \kappa|\lambda} | \hat{H}_t \Psi ~.
\label{eq41}
\end{align}
The action of $\hat{V}^{\lambda}_{t,\kappa}$ on $\tilde{\Psi}$ at the node $(\lambda)$
can be computed using a scheme analogous to Eq.(\ref{eq38}) with a numerical effort scaling
as $n \cdot N$ where $n$ denotes the number of SPFs associated with an edge and
$N$ the number of configurations associated with a node. Consequently, the numerical
effort associated with the computation of the
$\mathbf{X_{t}}^{\lambda \circ \kappa|\lambda}$ and the
$\hat{V}_{t;ij}^{\lambda \circ \kappa|\lambda}$ also scales as $n \cdot N$.
For an ideally structured tree, the revised non-hierarchical CDVR can thus be
implemented with no part of the calculation scaling worse than $n^4$.

The revised CDVR described above yields accurate results for separable potentials
(see Appendix B for a derivation). Thus, the optimal separable potentials discussed
in the context of the original non-hierarchical CDVR in Ref.\onlinecite{ElHoM2}
become obsolete in the revised CDVR.

\section{Quadrature optimized SPFs \label{sec5}}

Unoccupied SPFs do not contribute to the wavefunction representation and can thus
be redefined to increase the accuracy of the CDVR quadrature. A scheme to define
artifical SPFs that systematically increase the accuracy of the CDVR quadrature
has been introduced in Ref.\onlinecite{ElHoM}. A set of additional
SPFs at the edge $(\lambda \circ \kappa |\lambda)$ can be obtained by
diagonalizing the operator
\begin{align}
\hat{Y}^{\lambda \circ \kappa|\lambda} =
(1-\hat{P}^{\lambda \circ \kappa|\lambda})
\left(\sum_i \hat{x}_i \hat{\rho}^{\lambda \circ \kappa|\lambda}
\hat{x}_i \right)
(1-\hat{P}^{\lambda \circ \kappa|\lambda})
\label{eq42}
\end{align}
where the index $i$ numbers all physical coordinates $x_i$ present
$\phi^{\lambda \circ \kappa|\lambda}$ and
\begin{align}
\hat{\rho}^{\lambda \circ \kappa|\lambda} &= \sum_{j,l}
|\phi^{\lambda \circ \kappa|\lambda}_l \rangle
\rho^{\lambda \circ \kappa|\lambda}_{jl}
\langle \phi^{\lambda \circ \kappa|\lambda}_j | ~,
\label{eq43}
\\
\hat{P}^{\lambda \circ \kappa|\lambda} &=
\sum_{m} |\phi^{\lambda \circ \kappa|\lambda}_m \rangle
\langle \phi^{\lambda \circ \kappa|\lambda}_m | ~.
\label{eq44}
\end{align}
The eigenstates corresponding to largest eigenvalues of
$\hat{Y}^{\lambda \circ \kappa|\lambda}$ define artifical
SPFs that should be added to the SPF basis to improve the accuracy of the
CDVR quadrature. If the number of artifical SPFs resulting from
$\hat{Y}^{\lambda \circ \kappa|\lambda}$ is not sufficient, similiar
operators including higher powers of the $x_i$ can be devised \cite{ElHoM}.
For simplicity, these extensions will not be considered here explicitly.

Experience resulting from various test calculations showed that the use
of additional SPFs following this schemes negatively affects the time
steps used by the constant mean field (CMF) scheme \cite{MCTDHIntegrator,M4}
employed to integrate the MCTDH equations of motion. The artifical SPFs tend
to varying strongly with time forcing the integrator to update the matrix
elements extremely frequently. To address this issue, the modified scheme
described below has been devised.

In the modified scheme, the operator 
\begin{align}
\hat{B}^{\lambda \circ \kappa|\lambda} =
\hat{\rho}^{\lambda \circ \kappa|\lambda} + \epsilon_{unocc} \cdot
\left(\sum_i \hat{x}_i \hat{\rho}^{\lambda \circ \kappa|\lambda}
\hat{x}_i \right)
\label{eq45}
\end{align}
is diagonalized. Here $\epsilon_{unocc}$ defines the threeshold value when a
SPF is considered unoccupied. If $n$ SPFs are employed and all $n$
eigenvalues of $\hat{\rho}^{\lambda \circ \kappa|\lambda}$ are significantly
larger than $\epsilon_{unocc}$, the space spanned by the $n$ largest
eigenvalues of $\hat{B}^{\lambda \circ \kappa|\lambda}$ is identical to
the space spanned by the $n$ original SPFs. However, if $m$ SPFs show
populations significantly smaller than $\epsilon_{unocc}$ and $(n-m)$
SPFs show populations significantly larger than $\epsilon_{unocc}$, the
space spanned by the $n$ largest eigenvalues of
$\hat{B}^{\lambda \circ \kappa|\lambda}$ consists of the $(n-m)$ populated
SPFs and $m$ artifical SPFs similar to the eigenvectors corresponding
to the $m$ largest eigenvalues of $\hat{Y}^{\lambda \circ \kappa|\lambda}$.
In intermediate situations, eigenstates of
$\hat{B}^{\lambda \circ \kappa|\lambda}$ appear that are mixtures of SPFs
with populations close to $\epsilon_{unocc}$ and relevant eigenstates of
$\hat{Y}^{\lambda \circ \kappa|\lambda}$.

The eigenstates $|b^{\lambda,\kappa}_i \rangle$
corresponding to the $n$ largest eigenvalues $b^{\lambda,\kappa}_i$
of $\hat{B}^{\lambda \circ \kappa|\lambda}$ are then used
to obtain an improved set of $n$ new SPFs
$\phi^{\lambda \circ \kappa|\lambda}_i$. Specifically, first a
set of $n$ orthonormal functions
\begin{align}
|\chi^{\lambda,\kappa}_i \rangle &= \sum_{j=1}^n U^{\lambda,\kappa}_{i,j}
|b^{\lambda,\kappa}_j \rangle
\label{eq46}
\end{align}
is constructed where the unitary transformation matrix
$\mathbf U^{\lambda,\kappa}$ is chosen to maximize the overlap between
the old SPFs $\phi^{\lambda \circ \kappa|\lambda}_i$ and the
new $\chi^{\lambda,\kappa}_i$. Assigning a population
$b^{\lambda,\kappa}_i$ to each state $|b^{\lambda,\kappa}_i \rangle$,
the corresponding wavefunction reads (see Eq.(\ref{eq7}))
\begin{align}
|\Psi^{\lambda,\kappa} \rangle =
\sum_{i,j} U^{\lambda,\kappa}_{i,j} ~ \left(b^{\lambda,\kappa}_j
\right)^{\frac{1}{2}} |b^{\lambda,\kappa}_j 
\phi^{\lambda | \lambda \circ \kappa}_i \rangle ~.
\label{eq47}
\end{align}
Here the superscript $\lambda,\kappa$ indicates that the new wavefunction
is constructed using a new set of SPFs associated with the edge
$(\lambda \circ \kappa|\lambda)$.

The coefficients $\tilde{A}^\lambda_I$ of Eq.(\ref{eq10}) enter
in the definition of all $\phi^{\lambda \circ \kappa|\lambda}_i$
associated with any edge connected to the node $(\lambda)$.
However, different revised wavefunctions $|\Psi^{\lambda,\kappa} \rangle$
result from Eq.(\ref{eq47}) for different choices of $\kappa$.
To equally account for the effects of all edges, the new
$\tilde{A}^\lambda_I$ are computed using the average of all
$|\Psi^{\lambda,\kappa} \rangle$:
\begin{align}
\tilde{A}^\lambda_I &= \langle \Phi^\lambda_I | \Psi \rangle
= \frac{1}{\nu+1} \sum_{\kappa=0}^\nu \langle \Phi^\lambda_I |
\Psi^{\lambda,\kappa} \rangle ~.
\label{eq48}
\end{align}
Finally, the new $A^\lambda_{i_0,i_1,..,i_\nu}$ are obtained from
the new $\tilde{A}^\lambda_{i_0,i_1,..,i_\nu}$ by symmetric
reorthogonalization with respect to the index $i_0$.

As a result of replacing the $\tilde{A}^\lambda_I$ at all nodes
$(\lambda)$ accordingly, a revised wavefunction is obtained where
SPFs showing tiny populations are replace by artifical SPFs improving
the quadrature accuracy. Since these artifical SPFs show populations
of the order of $\epsilon_{unocc}$, the revised wavefunction slightly
differs for the original one. Thus, $\epsilon_{unocc}$ has to be chosen
sufficiently small to assure that relevant observables are not affected
by these populations. However, assigning these small populations to the artifical
SPFs guarantees their short time stability during the propagation and
successfully addresses the aforementioned problem with the short integration
steps. In all calculations employing this scheme presented in Sect.VI.C,
time steps comparable to the ones required in analogous standard MCTDH
calculations have been observed.

\section{Numerical examples \label{sec6}}

The properties of the revised non-hierarchical CDVR are illustrated
studying typical examples already considered in previous work on the
CDVR approach \cite{M3,vHM3,M6,M7,ElM5,ElHoM,ElHoM2}:
the photodissociation of NOCl, the vibrational states of the methyl radical,
and the S$_0$$\to$S$_2$ excitation in pyrazine.

\subsection{Photodissociation of NOCl \label{sec6b}}

The description of the photodissociation of NOCl closely follows
Ref. \onlinecite{MMC1}. A $NO+Cl$ Jacobi construction and the Jacobi
coordinates $r, R$ and $\theta$ are used.
The PES is derived from an ab initio potential developed by
Schinke et al. \cite{NOClFlaeche} and refitted to the form \cite{MMC1}
\begin{equation}
V(r, R, \theta) = \sum_{i} c_{i} \cdot v^{(R)}_i(R)
\cdot v^{(r)}_i(r) \cdot v^{(\theta)}_i(\theta).
\label{eq49}
\end{equation}
The initial wavefunction is a Gaussian wave packet modeling
the vibrational ground state of the S$_0$ potential energy surface.
Due to the SOP form of the PES, potential matrix elements can
efficiently be calculated either with or without the CDVR approach.
The same number of SPFs, $n$, is used in all degrees of freedom.  

\begin{figure}[ht]
\includegraphics[width=0.40\textwidth]{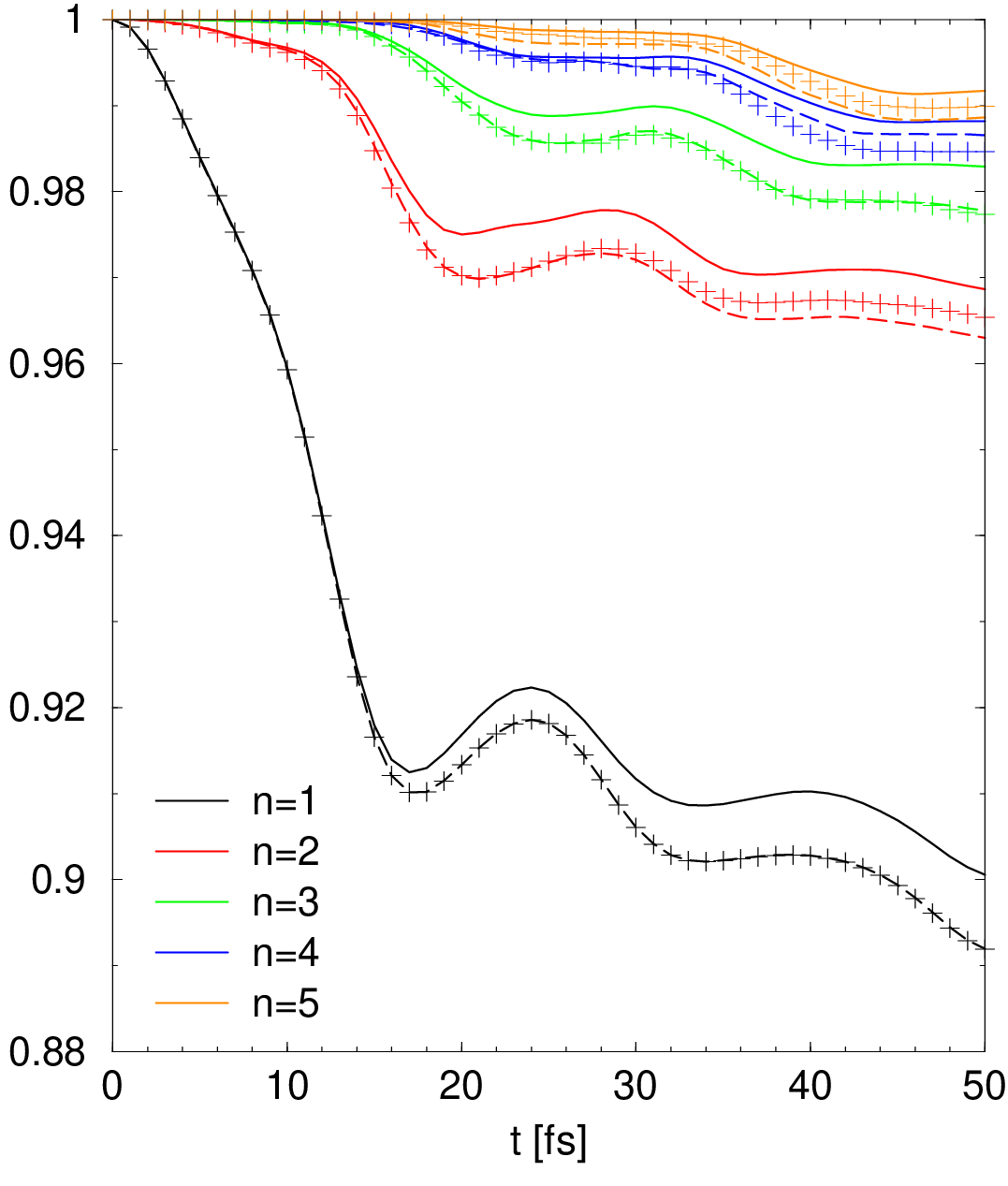}
\caption{Photodissociation of NOCl: real part of the overlap of a MCTDH
wavefunction and the numerically exact reference wavefunction. Results of
different types of MCTDH calculations are shown using different line styles:
solid lines for calculations using exact potential energy matrix elements,
dashed lines for calculations using the original non-hierarchical CDVR of
Refs.\onlinecite{ElM5,ElHoM2}, and plus signs for calculations using the
revised non-hierarchical CDVR. MCTDH calculations with different SPF
numbers $n$ are displayed using different colors as indicated in the legend.}
\end{figure}

In Fig.2, the accuracy of (single-layer) MCTDH calculations using revised
non-hierarchical CDVR approach is compared to the accuracy achieved
in MCTDH calculations using a SOP representation of the potential or
the original non-hierarchical CDVR \cite{ElM5,ElHoM2}.
Specifically, the real part of the overlap of a MCTDH wavefunction and the
numerically exact reference wavefunction is given. The real
of part of the overlap directly measures the accuracy the propagated MCTDH
wavefunction since
\mbox{$\|\psi_1 - \psi_2\| ^2$}$=$\mbox{$2[1-Re(<\psi_1|\psi_2>)]$}.
Different SPF basis set sizes ranging from Hartree level, $n=1$,
to effectively converged basis sets sizes, $n=5$, are considered.

Differences between the accuracy obtained with the original and
the revised CDVR scheme are minor. Depending on the number of SPFs
employed, the results computed with the revised scheme can be either
slightly more or slightly less accurate than the ones obtained with the
original non-hierarchical CDVR. In general, the inaccuracies resulting
from the use of either CDVR scheme are significantly smaller than errors
caused by the limited size of the SPF basis.

\subsection{Vibrational states of methyl}

The description of the methyl radical, CH$_3$, closely follows
Ref.\onlinecite{M5}. The coordinate system and quasi-exact kinetic
energy operator developed in Ref.\onlinecite{ENM} and the potential
energy surface of Medvedev et al. \cite{CH3PES} are used.
The coordinates $\rho$ and $\vartheta,\varphi$ describe the totally
symmetric and e-symmetric stretching modes, respectively.
The umbrella bending and e-symmetric bending motions are described
by the coordinates $\theta$ and coordinates $\phi,\chi$,respectively.
Block relaxation within the state-averaged (single-layer)
MCTDH approach \cite{M5} is used to compute the vibrational eigenstates.
The calculations employ the DVR schemes and number of grid points as described in
Ref. \onlinecite{M5}. The number of SPFs employed in the coordinates
$\rho$, $\theta$, $\vartheta$, $\varphi$, $\phi$, and $\chi$ is
4, 8, 8, 4, 6, and 6, respectively.   

\begin{table}
\caption{Zero point energy and vibrational excitation energies
E$_{\nu_1 \nu_2 \nu_3 \nu_4}$ of CH$_3$ (in cm$^{-1}$;
$\nu_1$, $\nu_2$, $\nu_3$, and $\nu_4$ label excitations in the totally
symmetric stretching, e-symmetric stretching, umbrella bending,
and e-symmetric bending modes, respectively). Results of MCTDH
calculations using the revised CDVR approach are compared with
results of original non-hierarchical CDVR \cite{ElM5} and
numerically exact results \cite{ENM}. Differences
$\Delta$E$_{\nu_1 \nu_2 \nu_3 \nu_4}$ between the MCTDH results
and the results of Ref. \onlinecite{ENM} indicate the error
resulting from the CDVR quadrature.\\}
\begin{tabular}{ c | r | r c | r c } 
\hline 
~ & Ref.\onlinecite{ENM}  & \multicolumn{2}{c|}{Ref.\onlinecite{ElM5}} & \multicolumn{2}{c}{revised}  \\[-0ex]
  & & \multicolumn{2}{c|}{(CDVR) } & \multicolumn{2}{c}{CDVR}\\  \hline
$(\nu_1 \nu_2 \nu_3 \nu_4)$ & E~~~~ & E~~~~& $\Delta$E&E~~~~& $\Delta$E\\ \hline
$(0000)$ 	&	~~6445.2&	~~6445.1&	0.1 &	~~6445.2&	0   	\\[-0ex]
$(0100)$	& 	591.5 	&	591.4 	&  	0.1 &	591.4	&	0.1   	\\[-0ex]
$(0200)$	& 	1265.7	&	1265.6	&	0.1 &	1265.7	&	0   	\\[-0ex]
$(0001)$	& 	1388.2	&	1388.2	&	0 &	1388.2	&	0   	\\[-0ex]
       		&	1388.5	&	1388.4	&	0.1 &	1388.5	&	0 	\\[-0ex]
$(0101^1)$	&	1991.4	&	1991.3	&	0.1 &	1991.3	&	0.1   	\\[-0ex]
		    &	1991.9	&	1991.9	&	0 &	1991.9	&	0   	\\[-0ex]
$(0300)$	&	1993.4	&	1993.3	&	0.1 &	1993.3	&	0.1   	\\[-0ex]
$(0201^1)$	&	2673.2	&	2673.2	&	0 &	2673.2	&	0 	\\[-0ex]
		    &	2673.3	&	2674.1	&	0.8 &	2674.1	&	0.8 	\\[-0ex]
$(0002^0)$	&	2750.7	&	2750.6	&	0.1 &	2750.7	&	0   	\\[-0ex]
$(0400)$	&	2762.3	&	2762.2	&	0.1 &	2762.3	&	0   	\\[-0ex]
$(0002^2)$	&	2767.6	&	2767.5	&	0.1 &	2767.6	&	0   	\\[-0ex]
		    &	2767.6	&	2767.6	&	0 &	2767.6	&	0   	\\[-0ex]
$(1000)$	&	2989.7	&	2989.0	&	0.7 &	2989.7	&	0 	\\[-0ex]
$(001^10)$	&	3142.3	&	3143.3	&	0.9 &	3141.8	&	0.5 	\\[-0ex]
		    &	3142.6	&	3143.9	&	1.3 &	3142.6	&	0 	\\[-0ex]
$(0102^0)$	&	3367.3	&	3367.2	&	0.1 &	3367.2	&	0.1 	\\[-0ex]
$(0102^2)$	&	3382.7	&	3382.6	&	0.1 &	3382.7	&	0 	\\[-0ex]
		    &	3382.8	&	3382.8	&	0 &	3382.8	&	0   	\\[-0ex]
$(0301^1)$	&	3406.6	&	3406.5	&	0.1 &	3406.5	&	0.1   	\\[-0ex]
		    &	3407.9	&	3407.7	&	0.2 &	3407.8	&	0.1   	\\[-0ex]
$(0500)$	&	3556.8	&	3556.0	&	0.8 &	3555.5	&	1.3   	\\[-0ex]
$(1100)$	&	3573.1	&	3572.0	&	1.1 &	3572.4	&	0.7 	\\[-0ex]
$(011^10)$	&	3709.3	&	3710.4	&	1.1 &	3708.8	&	0.5 	\\[-0ex]
		    &	3709.6	&	3711.0	&	1.4 &	3709.7	&	0.1 	\\[-0ex]
$(0202^0)$	&	4057.8	&	4057.6	&	0.2 &	4057.6	&	0.2 	\\[-0ex]
$(0202^2)$	&	4072.7	&	4072.6	&	0.1 &	4072.6	&	0.1 	\\[-0ex]
		    &	4073.0 	&	4072.9	&	0.1 &	4072.9	&	0.1   	\\[-0ex]
$(0003^1)$	&	4107.8	&	4107.6	&	0.2 &	4107.1	&	0.7 	\\[-0ex]
		    &	4108.5	&	4108.6	&	0.1 &	4108.0	&	0.5 	\\[-0ex]
$(0003^3)$	&	4137.5	&	4137.5	&	0 &	4137.6	&	0.1 	\\[-0ex]
$(0003^3)$	&	4138.3	&	4138.2	&	0.1 &	4138.0	&	0.3 	\\[-0ex]
$(0401^1)$	&	4178.3	&	4178.4	&	0.1 &	4178.3	&	0 	\\[-0ex]
		    &  	4179.9	&	4180.7	&	0.8 &	4179.9	&	0 	\\[-0ex]
$(1200)$	&	4234.7	&	4233.2	&	1.5 &	4233.6	&	1.1 	\\ \hline 
mean $\Delta$E  &           &           &   0.4 &           &   0.2     \\
\hline
\end{tabular}
\end{table}

Tab. I lists the first 36 vibrational levels computed with
MCTDH calculations employing the revised non-hierarchical CDVR introduced
in the present work. The computed energies are compared to
results of the original non-hierarchical CDVR taken from
Ref.\onlinecite{ElM5} and accurate reference results taken
from Ref.\onlinecite{ENM}. As already seen in the previous example
studied in Sect.\ref{sec6b}, the accuracy achieved with the revised
CDVR scheme is roughly equal to the accuracy of the original
non-hierarchical CDVR. On average, the results computed with the
revised scheme appear to be slightly more accurate than the ones
reported for the original scheme in Ref.\onlinecite{ElM5}. However,
given the numerical uncertainties resulting from the choice of the
various accuracy and regularization parameters employed in the
different calculations, these differences should not be considered
significant.

\subsection{The S$_0$$\to$S$_2$ excitation in pyrazine}

The $S_0 \to S_2$ excitation in pyrazine is a benchmark example of non-adiabatic 
dynamics on conically intersecting PESs.
\cite{Pyrazin3D1,Pyrazin3D2,Pyrazin4D1,Pyrazin4D2,MK2,PyrazinPI1,PyrazinPI2,WMC,WMC2,RWMC,DieterMLMCTDH,WestM2}
The dynamics following the $S_0 \to S_2$ excitation proceeds in steps 
(see, e.g., Ref.\onlinecite{MK2} for a detailed discussion): 
a compact wavepacket starts its motion on the upper ($S_2$) 
adiabatic potential energy surface (PES), reaches the conical intersection 
between the $S_2$ and $S_1$ surfaces, and undergoes an essentially complete
transition into the lower adiabatic ($S_1$) electronic state. In the subsequent
motion on the lower adiabatic PES, the wavepacket rapidly fragments
and visits almost the entire accessible phase space. The description of 
the different steps, which show very different dynamical character, provides a
challenging test for any quantum dynamics method. 

\begin{figure*}[ht]
\includegraphics[width=\textwidth]{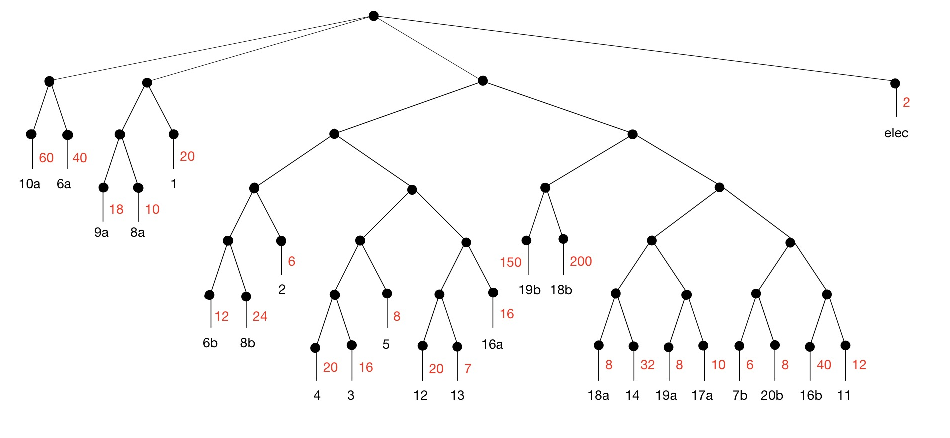}
\caption{Diagrammic representation of the multi-layer MCTDH wavefunction 
used to describe the $S_0 \to S_2$ excitation in pyrazine. The physical
degrees of freedom are labeled according to the vibration mode identifiers
used in Ref.\onlinecite{RWMC}. The sizes of the time-independent basis sets
are indicated by red numbers beside the corresponding edges. The numbers of
SPFs used in the various calculations are given in Tab.II.}
\end{figure*}

The full-dimensional (24D) quantum dynamic calculations employ the coupled
diabatic PESs developed by Raab~et~al.. \cite{RWMC} The Hamiltonian reads
\begin{align} 
\hat{H} = \sum_{n=1}^{24} \frac{\omega_n}{2} 
      \left( -\frac{\partial^2}{\partial q_n^2} +q_n^2 \right)
   + \left( \begin{array}{cc}
     V_1({\bf q})& V_c({\bf q})  \\
     V_c({\bf q}) & V_2({\bf q}) \\
   \end{array} \right)
\label{eq50}
\end{align}
with 
\begin{align}
V_i({\bf q}) = \sum_{n=1}^{24} a_n^{(i)} q_n 
+ \sum_{n=1}^{24} \sum_{m=n+1}^{24} b_{n,m}^{(i)} q_n q_m
\label{eq52}
\end{align}
(many of the $a_n^{(i)}$ and $b_{n,m}^{(i)}$ parameters vanish 
due to symmetry). It shows a sum of products (SOP) form and thus
can be readily be used in (multi-layer) MCTDH calculations without requiring CDVR.

The description used in the present work closely follows Ref.\onlinecite{ElHoM2}
where the quantum dynamics in S$_0$$\to$S$_2$ excited pyrazine was used
to study the accuracy of the non-hierarchical CDVR. The same tree structure
and time-independent basis sets are used in the multi-layer MCTDH wavefunction
representation (see Fig.3). As in Ref.\onlinecite{ElHoM2}, the Hamiltonian is written as 
\begin{align} 
\hat{H} =& \sum_{n=1}^{24} \frac{\omega_n}{2} 
      \left( -\frac{\partial^2}{\partial q_n^2} +q_n^2 \right)
     +V_1({\bf q}) \cdot \left( \begin{array}{cc}
     1 & 0  \\
     0 & 0 \\
   \end{array} \right)\nonumber\\
   &+ V_2({\bf q}) \cdot \left( \begin{array}{cc}
     0 & 0  \\
     0 & 1 \\
   \end{array} \right)
   + V_c({\bf q}) \cdot \left( \begin{array}{cc}
     0 & 1  \\
     1 & 0  \\
   \end{array} \right)
\label{eq53}
\end{align}
to obtain the form specified in Eq.(\ref{eq15}) for CDVR calculations.
Beside the SPF basis set used in Ref.\onlinecite{ElHoM2}, additional
larger basis set are considered. The use of larger basis sets in
MCTDH/CDVR calculations is facilitated by the increased numerical efficiency
of the revised CDVR scheme introduced in Sect.\ref{sec4}.
All SPF basis set employed are specified in Tab.II.

\begin{table}
\caption{Number of SPFs employed at each edge in the different multi-layer MCTDH 
calculations (for the definition of the tree used in the wavefunction
representation see Fig.3).\\}
\begin{tabular}{|l|c|c|c|c|c|c|c|} \hline
edge &  \multicolumn{7}{c|}{SPFs in basis} \\
     & B0 & B1 & B2 & B3 & B4 & B5 & R\\
\hline
1;1      & 10 & 20 & 32 & 41 & 49 & 57 & 80\\[-0.0ex]
2;11     & 11 & 13 & 15 & 18 & 22 & 26 & 50\\[-0.0ex]
2;12     &  5 &  6 &  7 & 11 & 13 & 15 & 40\\[-0.0ex]
1;2      &  6 &  7 & 10 & 13 & 16 & 19 & 40\\[-0.0ex]
2;21     &  4 &  6 &  6 &  8 & 10 & 12 & 40\\[-0.0ex]
3;211    &  4 &  5 &  4 &  6 &  7 &  8 & 18\\[-0.0ex]
3;212    &  3 &  3 &  3 &  4 &  5 &  6 & 10\\[-0.0ex]
2;22     &  4 &  4 &  5 &  6 &  7 &  8 & 20\\[-0.0ex]
1;3      &  9 & 17 & 28 & 36 & 43 & 49 & 80\\[-0.0ex]
2;31     &  5 &  6 & 15 & 19 & 23 & 27 & 50\\[-0.0ex]
3;311    &  3 &  5 &  7 & 11 & 13 & 15 & 50\\[-0.0ex]
4;3111   &  3 &  5 &  7 &  9 & 11 & 13 & 50\\[-0.0ex]
5;31111  &  2 &  4 &  4 &  5 &  6 &  7 & 12\\[-0.0ex]
5;31112  &  3 &  3 &  4 &  5 &  6 &  7 & 24\\[-0.0ex]
4;3112   &  2 &  2 &  2 &  3 &  4 &  5 &  6\\[-0.0ex]
3;312    &  4 &  9 & 14 & 20 & 24 & 28 & 50\\[-0.0ex]
4;3121   &  3 &  6 &  9 & 12 & 14 & 16 & 50\\[-0.0ex]
5;31211  &  3 &  5 &  7 &  9 & 11 & 13 & 50\\[-0.0ex]
6;312111 &  2 &  4 &  5 &  4 &  5 &  6 & 20\\[-0.0ex]
6;312112 &  3 &  3 &  4 &  4 &  5 &  6 & 16\\[-0.0ex]
5;31212  &  3 &  3 &  4 &  4 &  5 &  6 &  8\\[-0.0ex]
4;3122   &  3 &  5 &  8 & 11 & 13 & 15 & 50\\[-0.0ex]
5;31221  &  3 &  4 &  6 &  7 &  8 &  9 & 50\\[-0.0ex]
6;312211 &  3 &  4 &  4 &  6 &  7 &  8 & 20\\[-0.0ex]
6;312212 &  2 &  2 &  3 &  3 &  4 &  5 &  7\\[-0.0ex]
5;31222  &  3 &  4 &  4 &  5 &  6 &  7 & 16\\[-0.0ex]
2;32     &  8 & 14 & 26 & 33 & 40 & 47 & 80\\[-0.0ex]
3;321    &  7 & 13 & 20 & 15 & 18 & 21 & 50\\[-0.0ex]
4;3211   &  4 & 11 & 17 & 20 & 24 & 28 & 50\\[-0.0ex]
4;3212   &  5 & 12 & 19 & 23 & 28 & 33 & 50\\[-0.0ex]
3;322    &  6 & 14 & 24 & 31 & 37 & 43 & 80\\[-0.0ex]
4;3221   &  5 &  7 & 11 & 11 & 13 & 15 & 50\\[-0.0ex]
5;32211  &  5 &  7 &  9 & 11 & 13 & 15 & 50\\[-0.0ex]
6;322111 &  2 &  2 &  3 &  4 &  5 &  6 &  8\\[-0.0ex]
6;322112 &  5 &  7 &  8 & 10 & 12 & 14 & 32\\[-0.0ex]
5;32212  &  2 &  3 &  4 &  8 & 10 & 12 & 50\\[-0.0ex]
6;322121 &  2 &  2 &  3 &  3 &  4 &  5 &  8\\[-0.0ex]
6;322122 &  2 &  3 &  4 &  4 &  5 &  6 & 10\\[-0.0ex]
4;3222   &  4 &  8 & 11 & 14 & 17 & 20 & 50\\[-0.0ex]
5;32221  &  2 &  3 &  3 &  4 &  5 &  6 & 42\\[-0.0ex]
6;322211 &  2 &  2 &  3 &  3 &  4 &  5 &  6\\[-0.0ex]
6;322212 &  2 &  3 &  3 &  4 &  5 &  6 &  8\\[-0.0ex]
5;32222  &  5 &  7 &  9 & 10 & 12 & 14 & 50\\[-0.0ex]
6;322221 &  3 &  4 &  5 &  6 &  7 &  8 & 40\\[-0.0ex]
6;322222 &  4 &  5 &  5 &  6 &  7 &  8 & 12\\[-0.0ex]
1;4      &  2 &  2 &  2 &  2 &  2 &  2 &  2\\
\hline
\end{tabular}
\end{table}

Basis set B0 to B4 have already been used in Ref.\onlinecite{ElHoM2}.
The new basis set B5 was obtained by increasing the number of all SPFs
by about 15 percent compared to basis B4. This increases the total number
of A-coefficients and CDVR grid points by a factor of about 1.5 compared
to basis set B4. An even larger basis R was designed to provide
reference wavefunctions of comparatively high accuracy.

\begin{figure*}[t]
\includegraphics[width=0.40\textwidth]{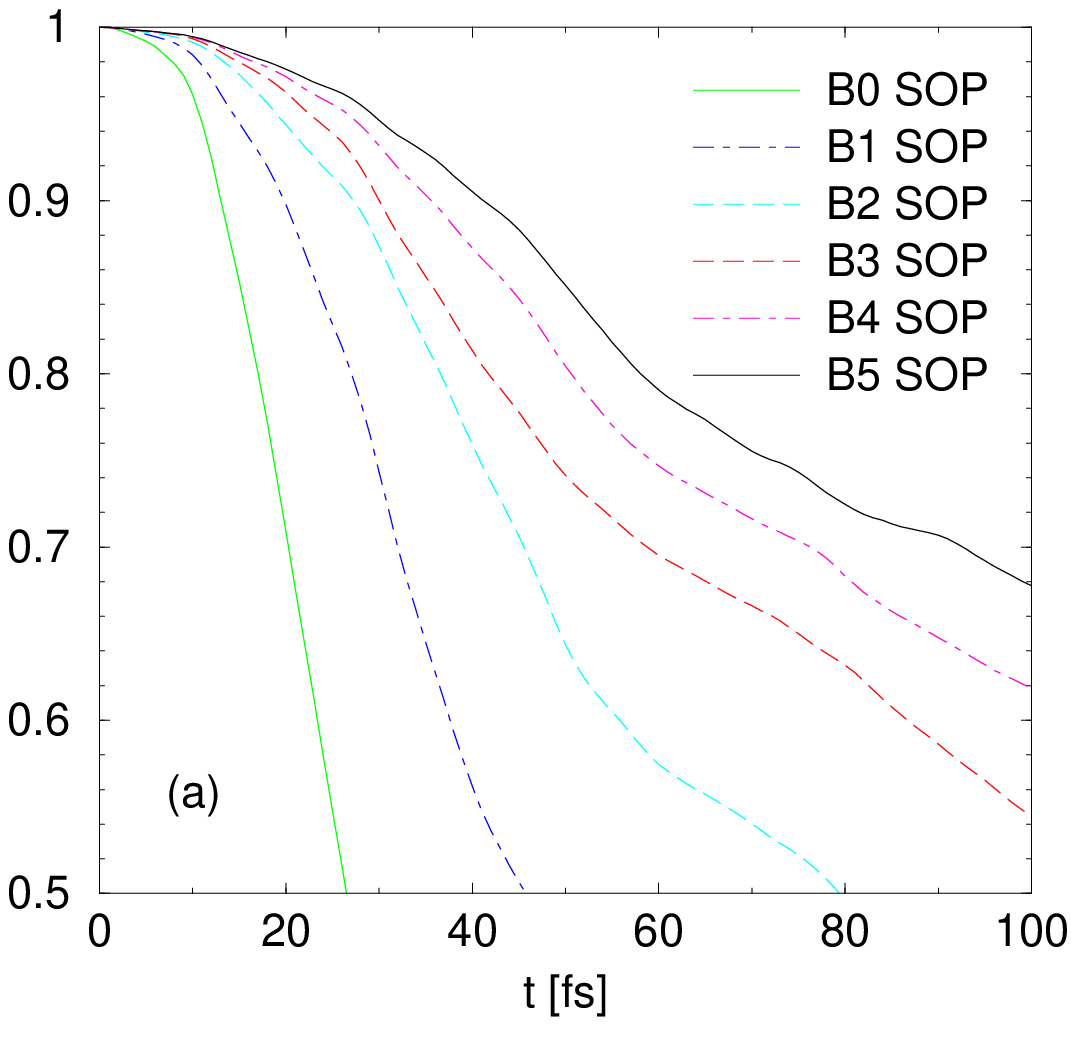}
\hspace{0.01\textwidth}
\includegraphics[width=0.40\textwidth]{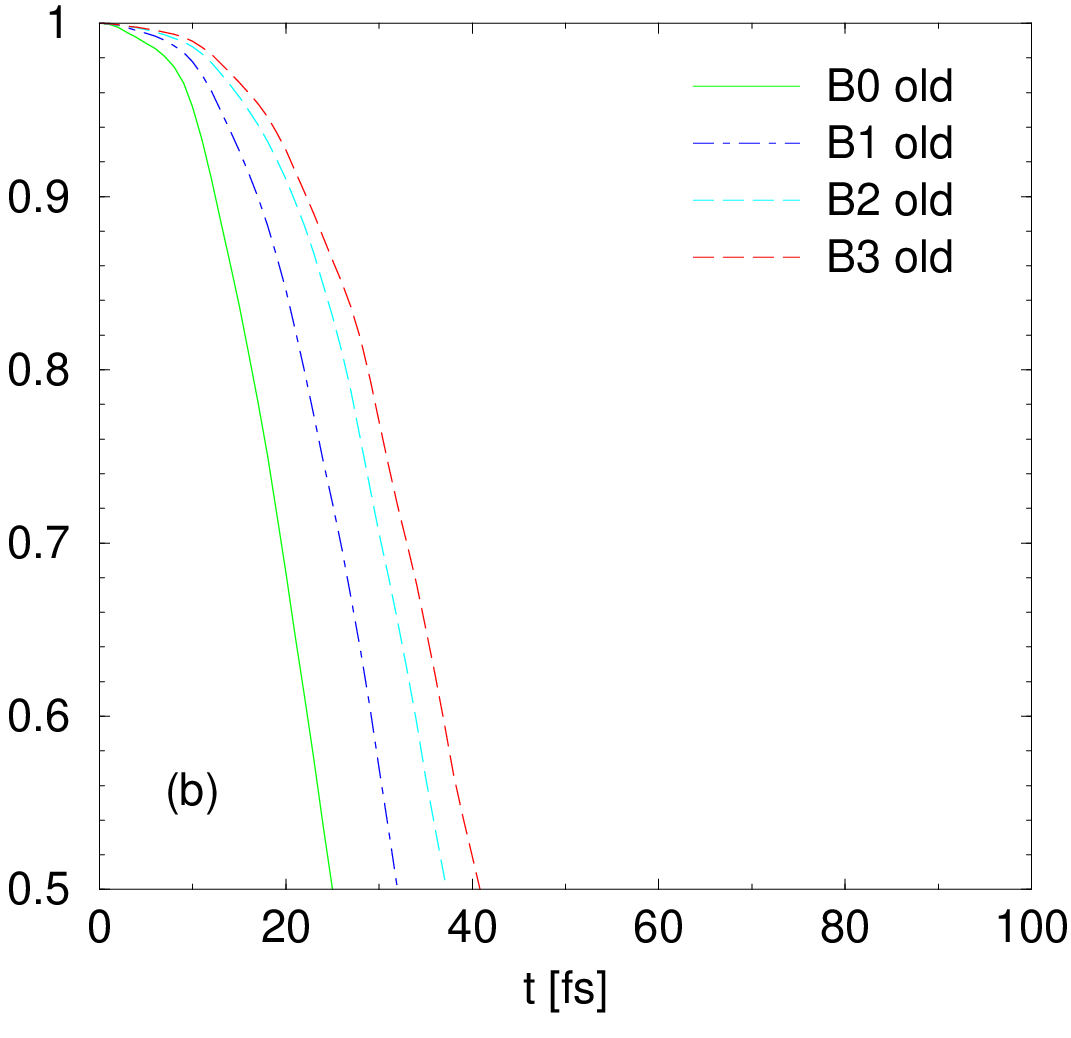}
\hspace{0.01\textwidth}\\
\includegraphics[width=0.40\textwidth]{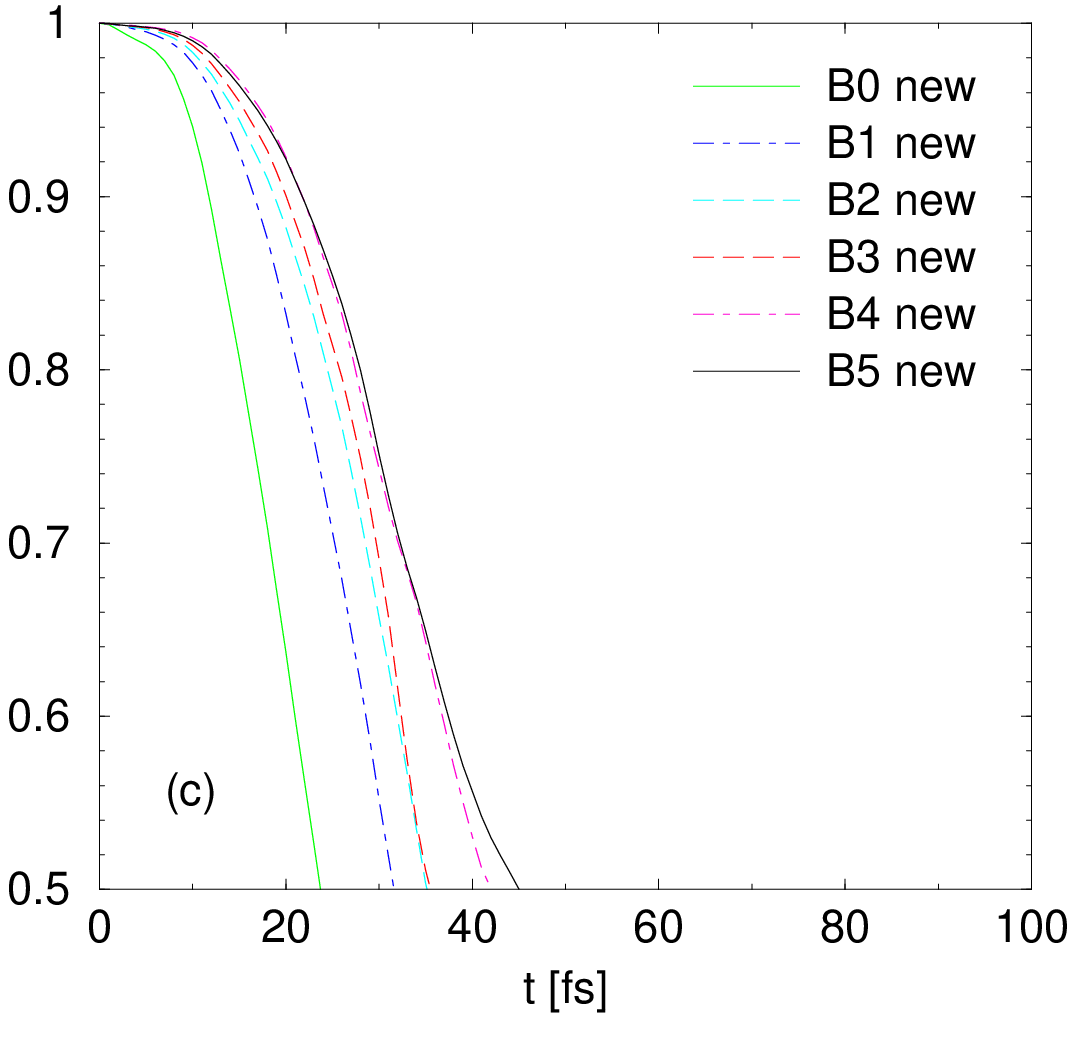}
\hspace{0.01\textwidth}
\includegraphics[width=0.40\textwidth]{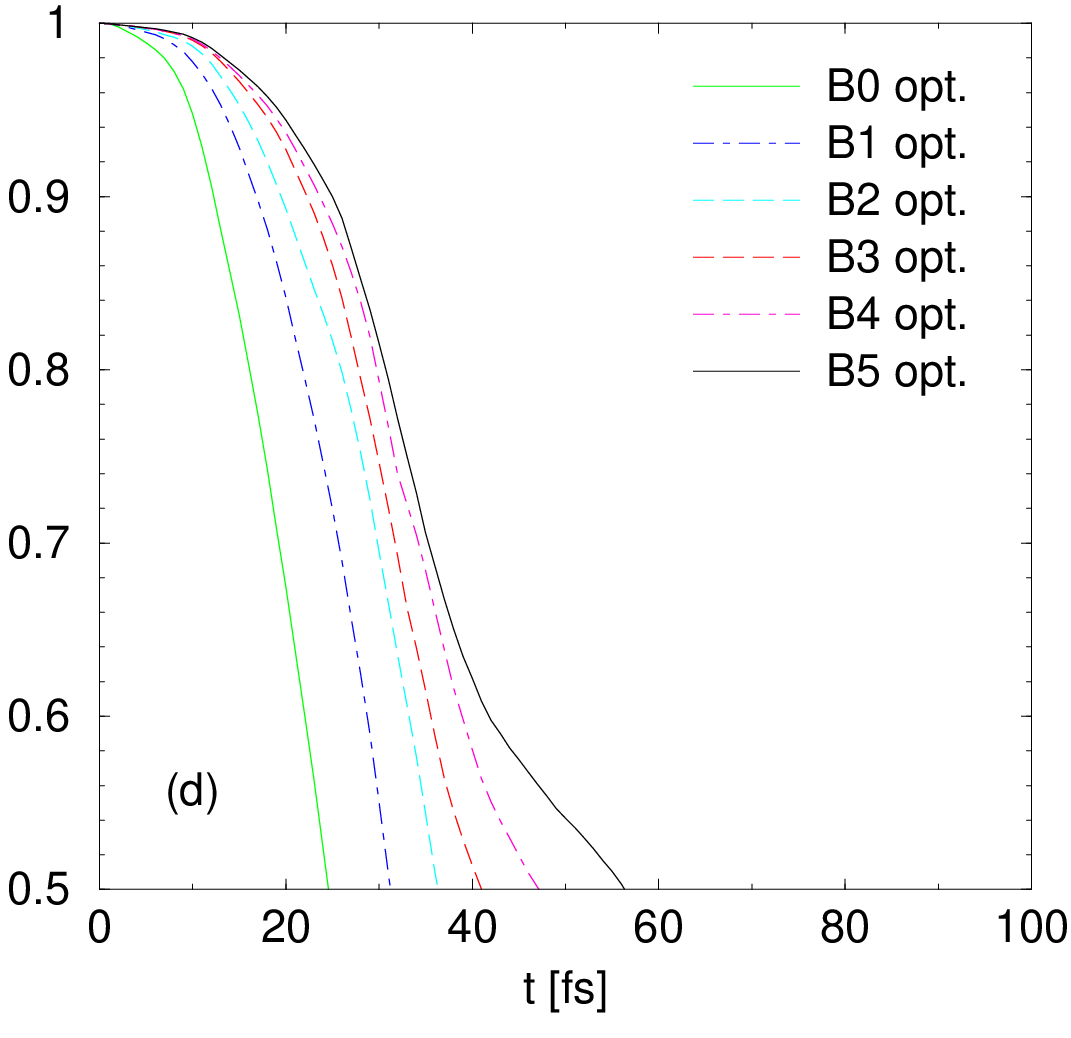}
\caption{Real part of the overlap $\langle \psi_{R,SOP}(t) | \psi(t) \rangle$ 
between the multi-layer MCTDH wavefunctions $ \psi(t)$ obtained by different calculations 
with reference results obtained with the basis R and a SOP representation of
the Hamiltonian:
Panels (a) to (d) show results of multi-layer MCTDH calculations using
a SOP representation, the original non-hierarchical CDVR approach \cite{ElHoM2},
the revised non-hierarchical CDVR approach of the present work, and
the revised non-hierarchical CDVR approach with optimized unoccupied SPFs
(see text for details), respectively. Results obtained with different SPF
basis sets size as indicated in the legend are displayed.}
\end{figure*}

\begin{figure*}[ht]
\includegraphics[width=0.40\textwidth]{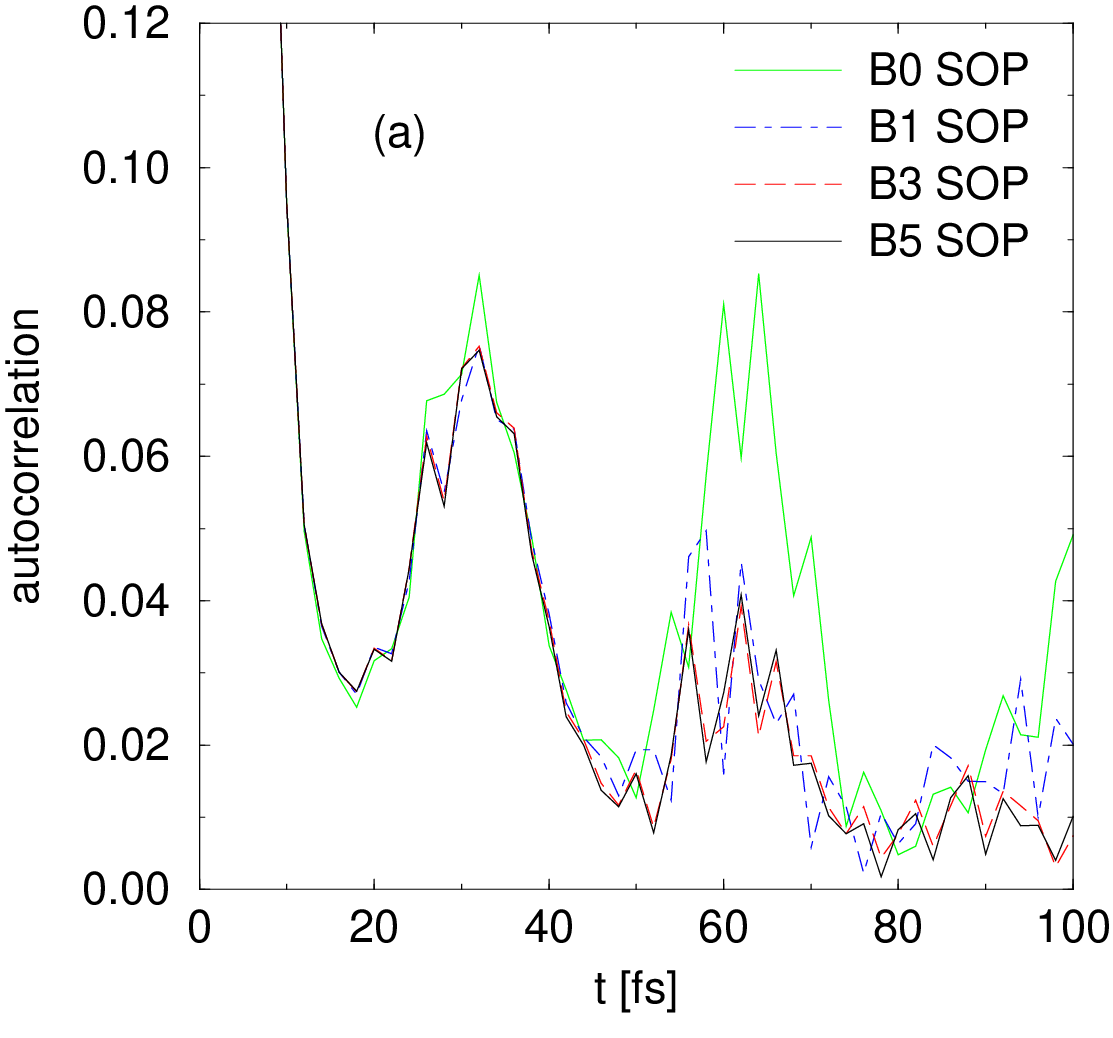}
\hspace{0.01\textwidth}
\includegraphics[width=0.40\textwidth]{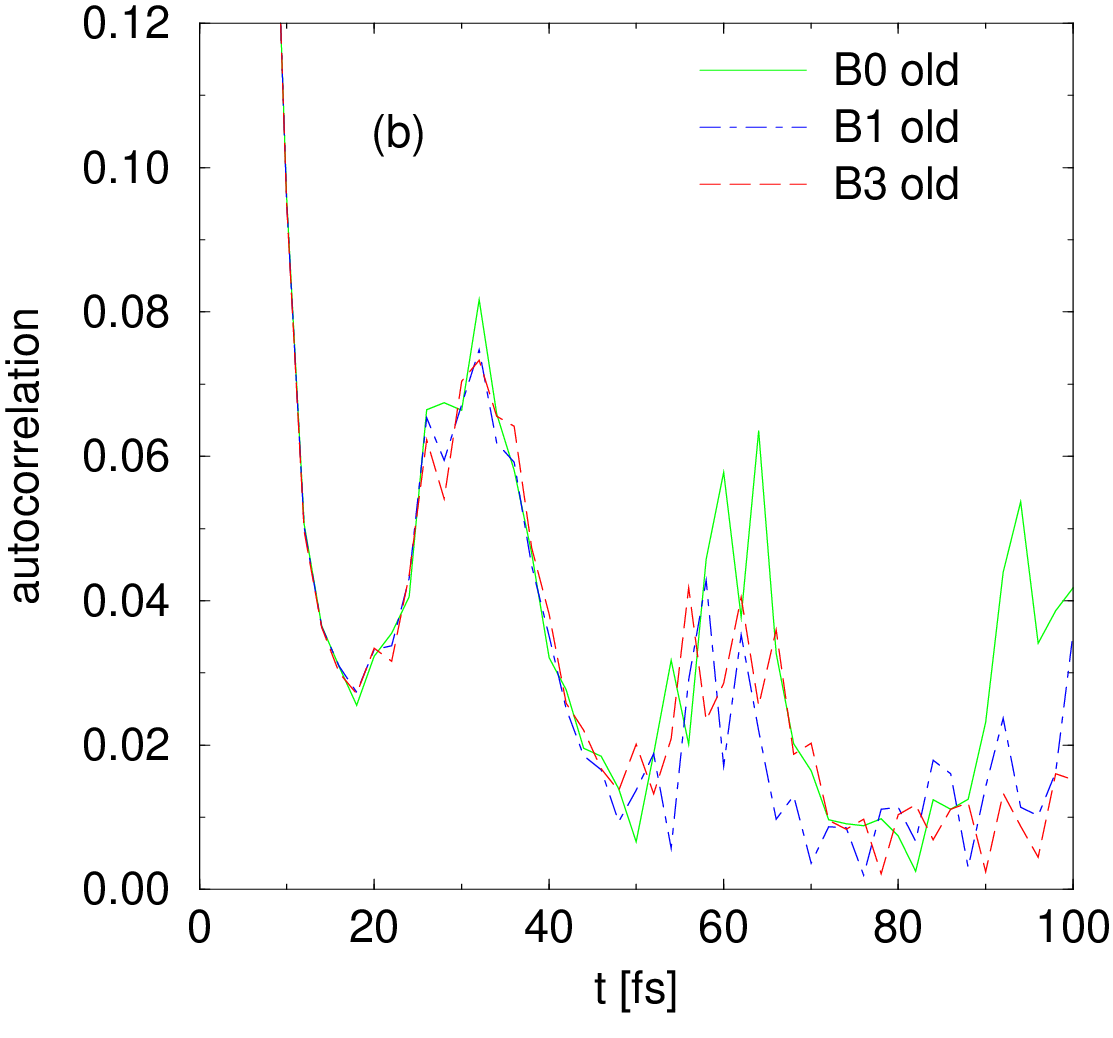}
\hspace{0.01\textwidth}\\
\includegraphics[width=0.40\textwidth]{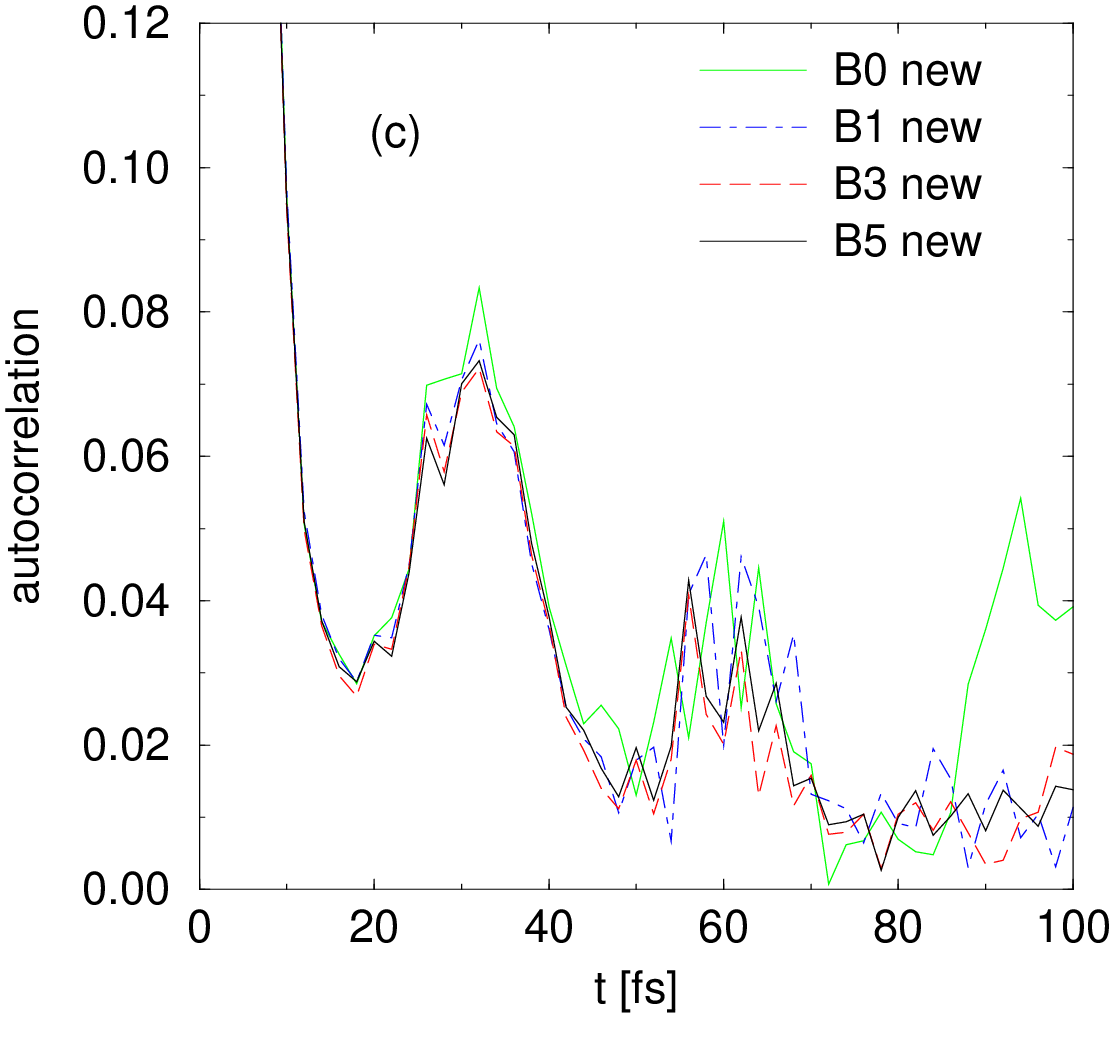}
\hspace{0.01\textwidth}
\includegraphics[width=0.40\textwidth]{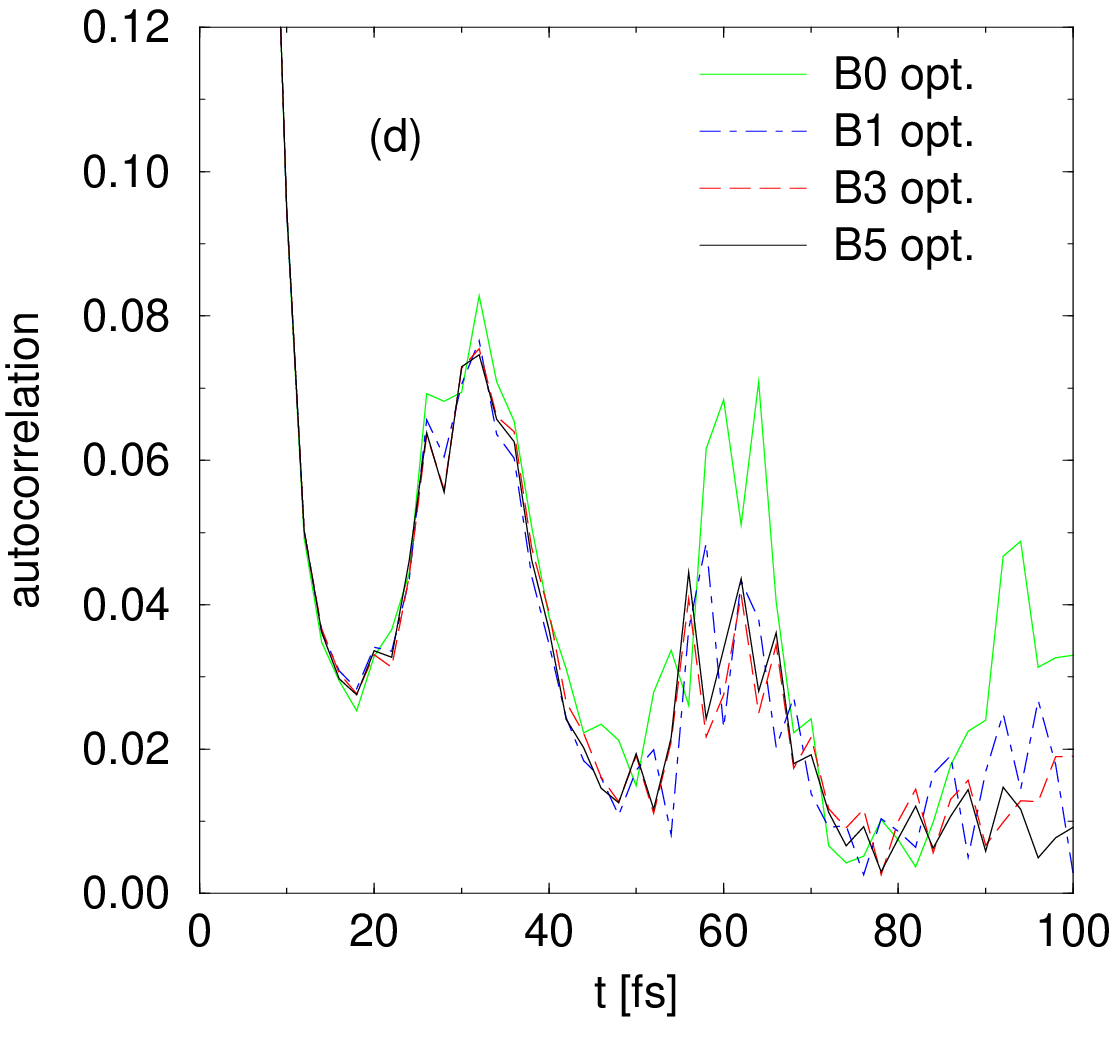}
\caption{The absolute values of the autocorrelation function 
obtained by different multi-layer MCTDH calculations are displayed as a function of time.
Panels (a) to (d) show results of multi-layer MCTDH calculations using a SOP
representation, the original non-hierarchical CDVR approach \cite{ElM5,ElHoM2},
the revised non-hierarchical CDVR approach of the present work, and
the revised non-hierarchical CDVR approach with optimized unoccupied SPFs
(see text for details), respectively. Results obtained with different SPF
basis sets size as indicated in the legend are displayed.}
\end{figure*}

\begin{figure*}[t]
\includegraphics[width=0.40\textwidth]{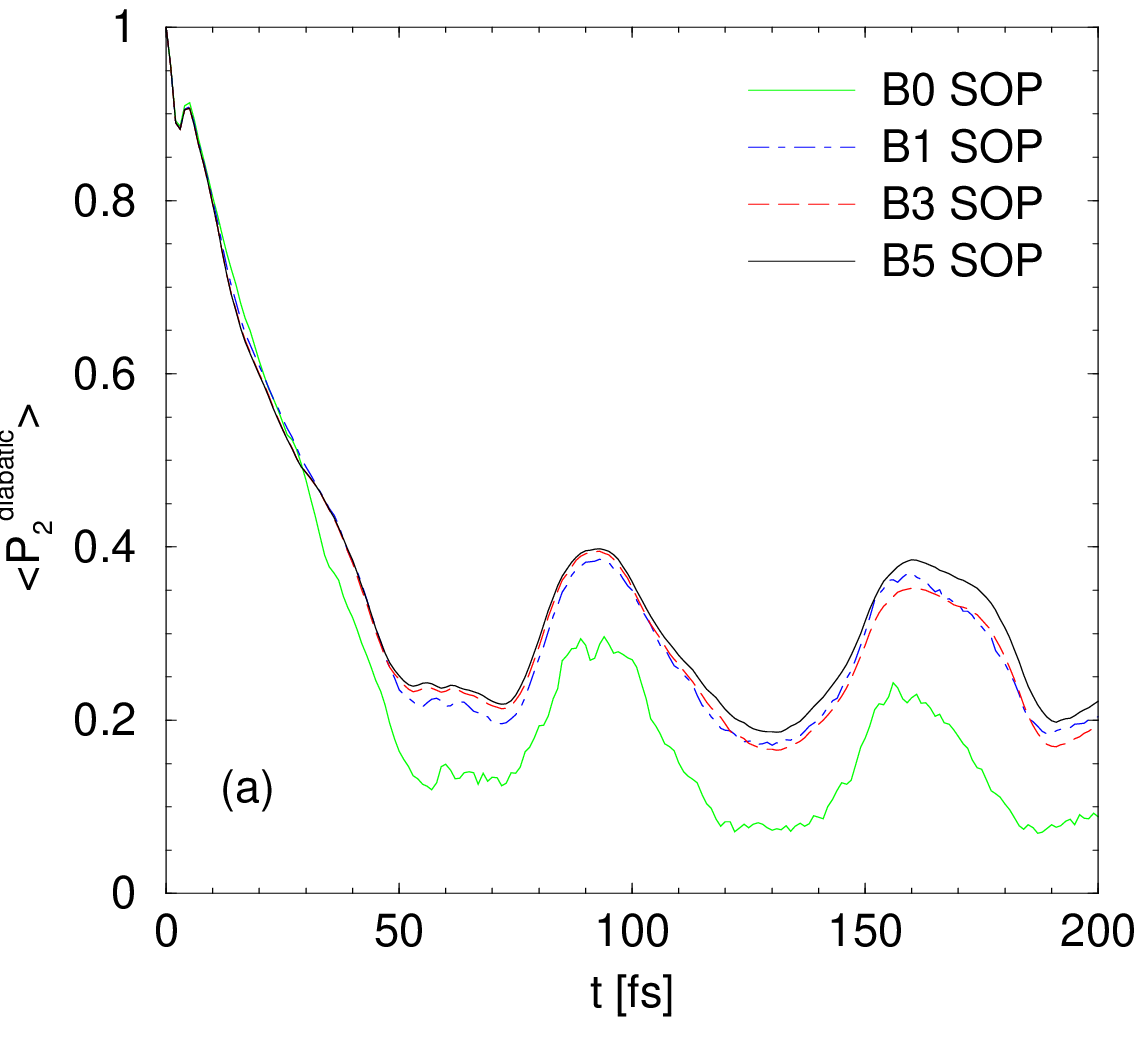}
\hspace{0.01\textwidth}
\includegraphics[width=0.40\textwidth]{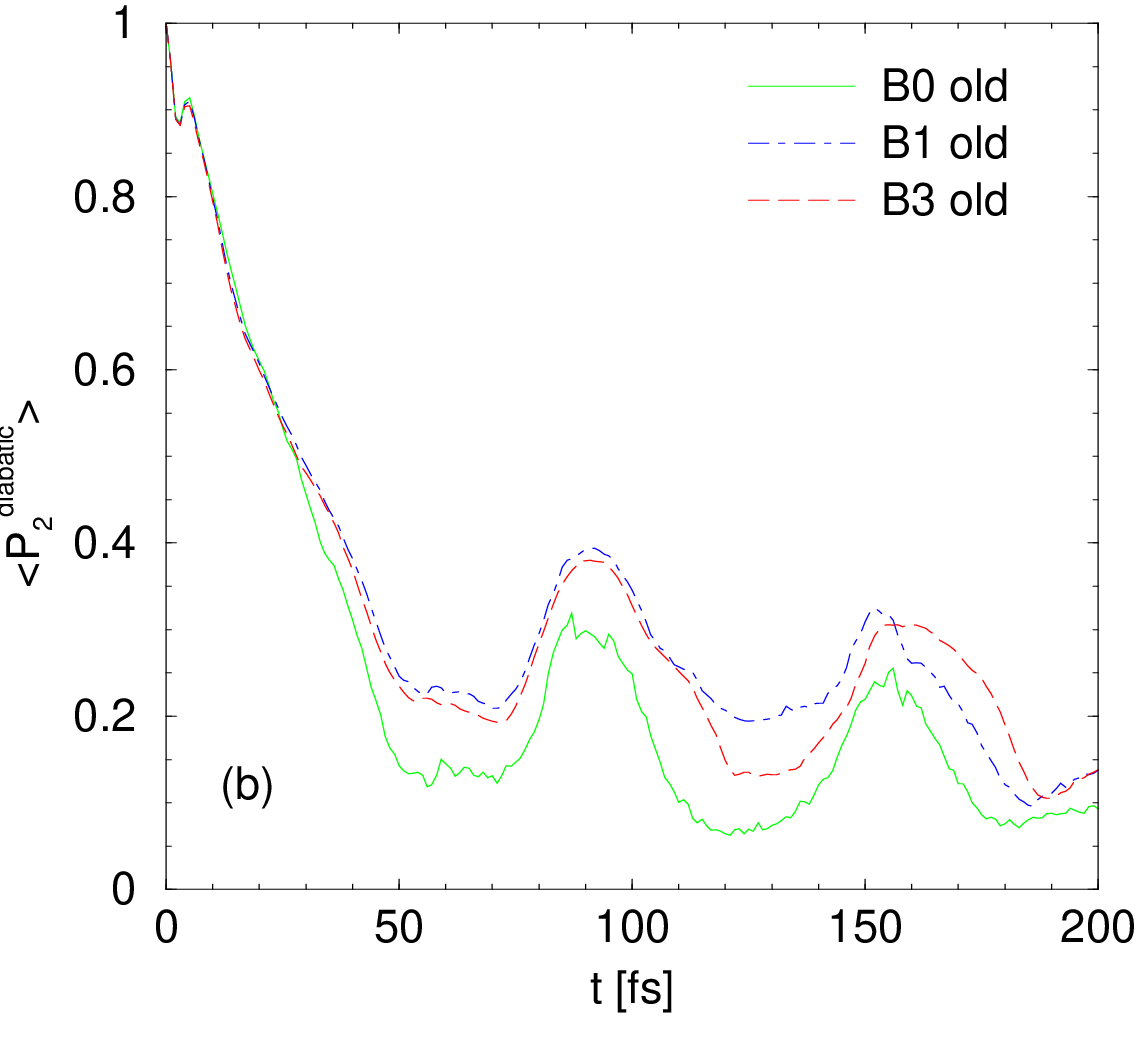}
\hspace{0.01\textwidth}\\
\includegraphics[width=0.40\textwidth]{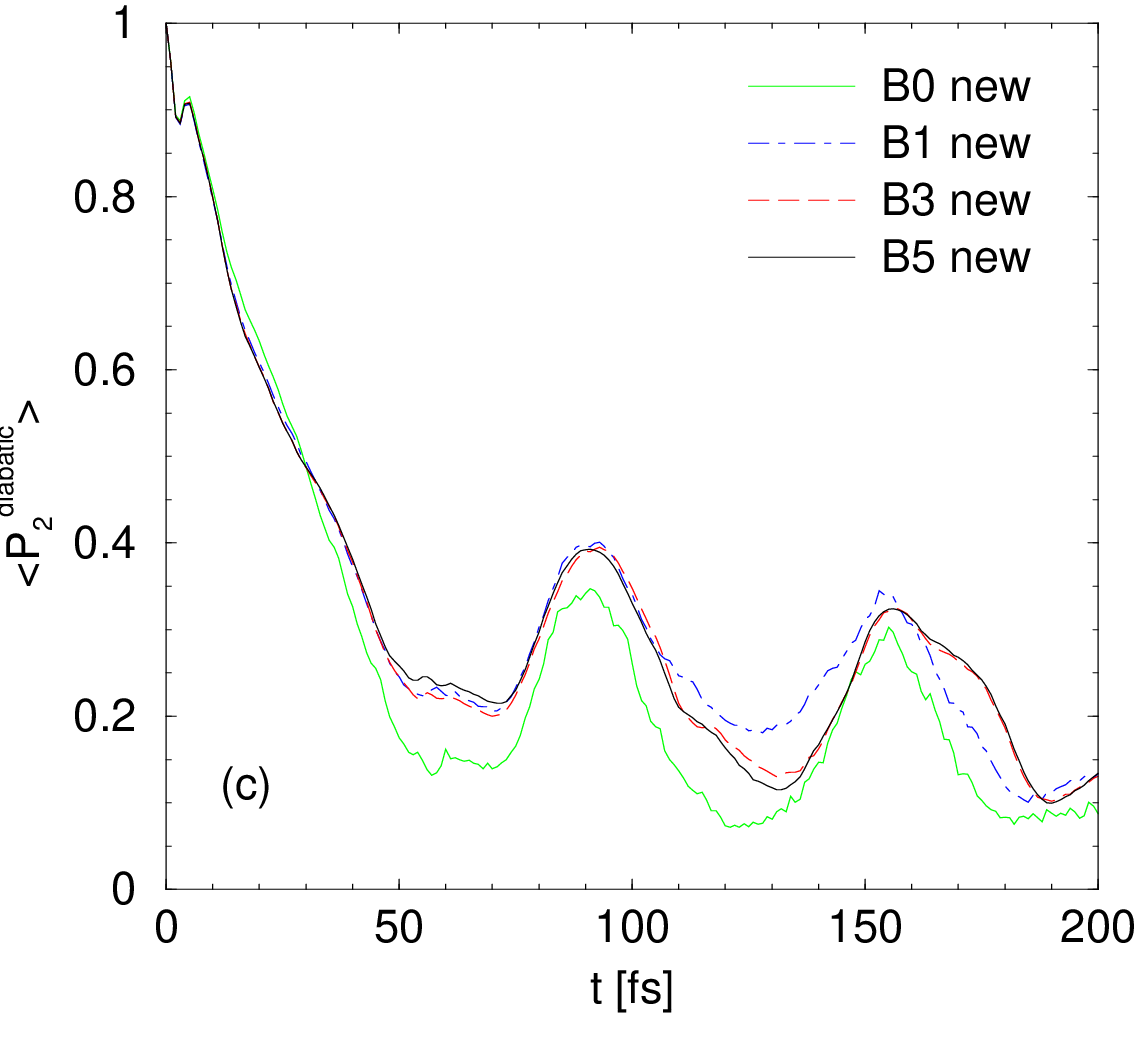}
\hspace{0.01\textwidth}
\includegraphics[width=0.40\textwidth]{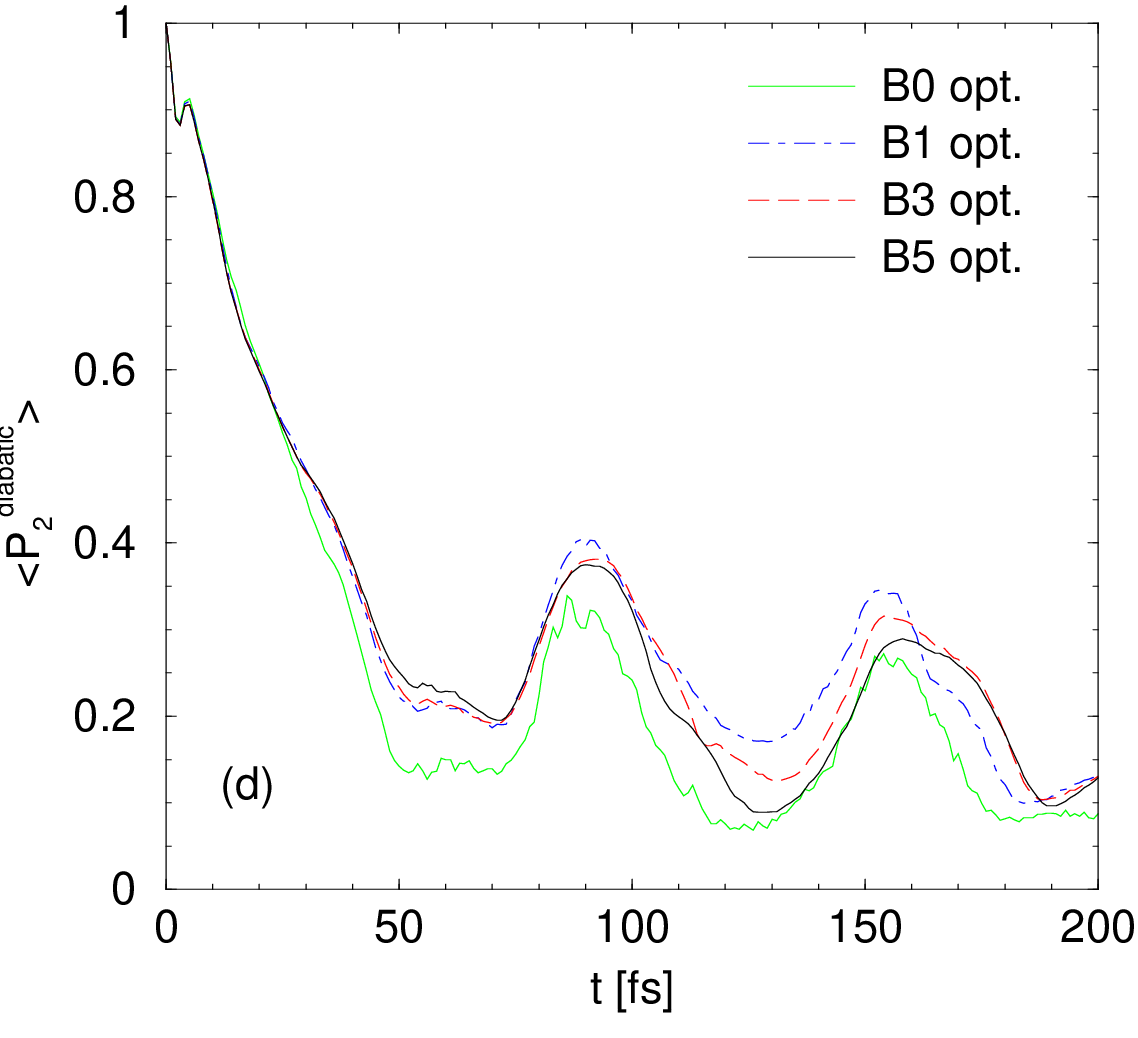}
\caption{The population of the second diabatic electronic state computed 
by different multi-layer MCTDH calculations with different basis sets (indicated in the
legend) is displayed as a function of time.
Panels (a) to (d) show results of multi-layer MCTDH calculations using a SOP
representation, the original non-hierarchical CDVR approach\cite{ElM5,ElHoM2},
the revised non-hierarchical CDVR approach of the present work, and
the revised non-hierarchical CDVR approach with optimized unoccupied SPFs
(see text for details), respectively. Results obtained with different SPF
basis sets size as indicated in the legend are displayed.}
\end{figure*}

In Figs. 4 to 6, results obtained by multi-layer MCTDH calculations
using different non-hierarchical CDVR schemes are compared to the
results of MCTDH calculations employing exact potential energy matrix
elements (i.e., MCTDH calculations using a SOP representation of the
entire Hamiltonian). In each figure,
the results of MCTDH calculations using the SOP representation are shown
in panel (a) and the results computed in with original non-hierarchical
CDVR taken from Refs.\onlinecite{ElHoM2} are displayed in panel (b).
Panel (c) shows results obtained with revised non-hierarchical
CDVR approach introduced in the present work. Results using the
revised non-hierarchical CDVR approach and the quadrature optimized
SPFs described in Sect.\ref{sec5} are displayed in panel (d). In
the calculations with the quadrature optimized SPFs, the threeshold
value $\epsilon_{unocc}$ was taken equal to the general accuracy
parameter $\epsilon$ used in the integration of the
SPF equations of motion and the regularisation of the
single-particle density matrices: $\epsilon_{unocc}=\epsilon=10^{-8}$.

In Fig.4, the accuracy of multi-layer MCTDH wavefunctions computed using different
schemes and SPF basis set sizes is compared. The real part of the
overlap of a MCTDH wavefunction and the reference wavefunction computed
employing the reference basis R and the SOP representation of the Hamiltonian
are shown. Since
\mbox{$\|\psi - \psi_{R,SOP}\| ^2$}$=$\mbox{$2[1-Re(<\psi_{R,SOP}|\psi>)]$},
the real of part of the overlap directly measures the accuracy the propagated
MCTDH wavefunction.

As already discussed in the previous work on the original non-hierarchical
CDVR, the relative accuracy achieved with the SOP and CDVR approaches vary
strongly depending on the size of the SPF basis. For the small B0 basis,
the differences between the SOP and any of the CDVR results are minor. Here
the additional error introduced by the CDVR quadrature is small compared to
the error in the wavefunction representation resulting for the small SPF basis.
Considering larger SPF bases, the relative importance of the CDVR quadrature
errors rises quickly with increasing basis set size.
This difference can be attributed to the change in the nature of the 
underlying dynamical processes. Initially, the wavepacket moves on the
upper adiabatic PES. It shows a compact form  and the CDVR approach can
easily achieve an accurate quadrature. After 10 and 20~fs, the wavepacket
passes from the upper to the lower adiabatic PES and starts fragmenting.
Achieving an accurate quadrature for the increasingly delocalized and
fragmented wavepacket is extremely demanding. Consequently, increasingly
large differences between the SOP and CDVR results appear for SPF bases
that cover the complex structure developing in the wavefunction in course
of time.

Comparing the results computed with the original and the revised
non-hierarchical CDVR displayed in panels (b) and (c) of Fig.4,
only minor differences can be observed. The results obtained by the
two approaches are essentially indistinguishable for the small
basis sets B0 and B1. For the largest basis set shown in Fig.4b,
B3, the differences become notable. Here the calculations
with the the original non-hierarchical CDVR yield slightly
more accurate results. Results computed with the revised CDVR and the
larger basis sets B4 and B5 are more accurate than the best results
shown in panel (b). Due to the improved scaling of the revised CDVR,
these calculations are numerically less demanding than calculations
with the original non-hierarchical CDVR and basis set B3.

Improvements can be achieved if the unoccupied SPFs are replace
by quadrature optimized SPFs as described in Sect.\ref{sec5}.
The results of these calculations are displayed in Fig.4d. For
the small basis set B0 and B1, the results are essentially identical
to the one displayed in panels (b) and (c). A relevant improvement
compared to the calculations without quadrature optimized SPFs can
be seen for the larger basis set B3, B4, and B5. For basis set
B3, the calculation using the revised CDVR and quadrature optimized
SPFs are as accurate as the calculations with the original
non-hierarchical CDVR.

As seen in Fig.4, it is extremely hard to achieve an accurate
description of the entire wavefunction over the entire time scale
considered. Once the dominant fraction of the wavepacket has passed to
the lower adiabatic PES after 20 to 30 ~fs, the wavepacket displays
chaotic-type dynamics and quickly fragments. Describing all the tiny
emerging fragments correctly is an effectively impossible task even
for SOP calculations. Furthermore, errors resulting from an unprecise
CDVR quadrature quickly increase due to the chaotic nature of the
underlying dynamics. However, relevant physical observables are
typically averaged quantities than can be reliably computed even
in presence of chaotic dynamics. For photoexcited pyrazin,
converged result can be obtained for many important physical
observables as, e.g., the absorption spectrum or the population
of the diabatic and adiabatic electronic states.

The autocorrelation function enables a detailed investigation of
the various features appearing in the absorption spectrum.
The absolute values of autocorrelation functions computed
with the different approaches are shown in Fig.5.
For any basis except the small B0 one, essentially converged results
for times up to 50~fs are obtained using either the SOP or any of the
CVDR schemes. Only for $t>50~fs$, appreciable difference between the
CDVR and SOP calculations start to appear. Comparing the results
of different CVR schemes shown in panels (b) to (d), no relevant
difference in the quality of the results can be observed.

The population of the second diabatic electronic state is
investigated in Fig.6. Calculations utilizing the SOP form
of the Hamiltonian achieve essentially converged results over
the entire time interval for all but the smallest SPF bases
(see panel (a)). All CDVR calculations provide similarly
accurate results for times up to about about 100~fs. For
later times, the CDVR calculations are less accurate 
but provide a sound description of the population dynamics.
Again, no relevant difference in the quality of the results
obtained by the different CDVR schemes can be observed (see
panel (b) to (d)).

\begin{figure}[t]
\includegraphics[width=0.40\textwidth]{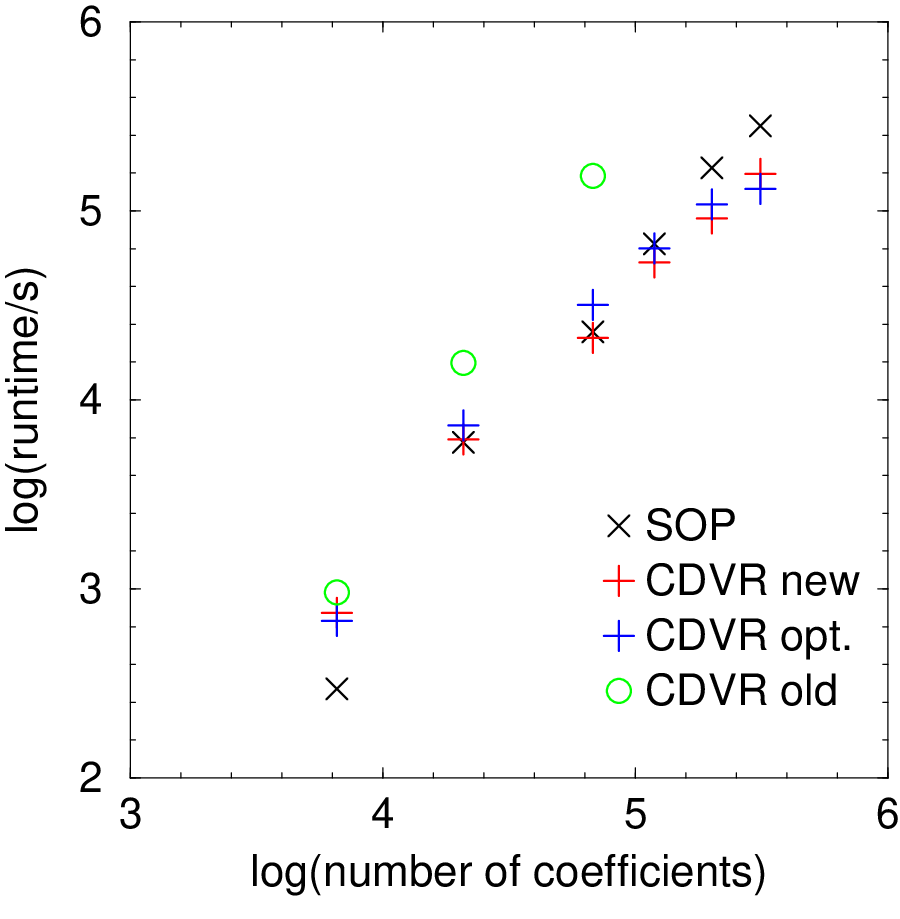}
\caption{Wall clock timings for calculations using the SOP
scheme, the revised CDVR (CDVR new), the revised CDVR
with quadrature optimized SPFs (CDVR opt.), and the original
non-hierarchical CDVR (CDVR old) with different
basis set sizes: the logarithm of the CPU time required
is plotted versus the logarithm of the total number of
A-coefficients used in the wavefunction representation.
Results for basis set B0, B1, B2, B3, B4, and B5 are given.
Note that in the figure a function $y=const.\cdot x^{4/3}$
would correspond to a line with a 45 degree slope.}
\end{figure}

An central advantage of the revised CDVR approach is the
improved scaling with the numerical effort. As discussed in
Sect.\ref{sec4}, the CPU time required by calculations
using the revised CDVR, $n$ SPF, and $N$ configurations
at a node ideally scales as $n \cdot N$. Multi-layer MCTDH calculations
using a SOP-Hamiltonian show the same scaling of the numerical
effort. For optimally structured tree where each node has
three edges, $N=n^3$. Thus, the required CPU time should
ideally scale as $n^4=N^{4/3}$. However, indirect effects
of the SPF basis set size, e.g., the dependence of the step
sizes required in the integration scheme on the number of
SPFs employed, can also affect the CPU time require and
modify the scaling.

In Fig.7, the CPU times required by the different calculations
considered above are analyzed. Results for multi-layer MCTDH calculations
with the basis sets B0, B1, B2, B3, B4, and B5 using
the SOP form, the revised CDVR, and the revised CDVR
and quadrature optimized SOPs are compared. Disregarding
differences between the specific numbers of SPFs used at
different edges, one would expect the CPU times should roughly
scale as \mbox{(\it total number of A coefficients\rm)$^{4/3}$}. The results
shown in Fig.7 confirm this expectation for all types
of calculations considered. Calculations using the revised
CDVR do not scale worse than calculations utilizing the SOP form.
(Actually, the revised CDVR calculation show a slightly more
favorable scaling resulting from a slightly difference dependence
of the step size in the integration on the SPF basis set size.)

Moreover, the absolute CPU times required by calculations using
the SOP form or the revised CDVR approches for the same basis set
are quite similar. Thus, the computational costs of the revised
CDVR calculations are comparable to the SOP calculations.
Considering that the pyrazine system studied is ideally suited for
MCTDH calculations using the SOP form, this result is remarkable and
impressively demonstrates the numerical efficiency achieved by the
revised CDVR approach.

Furthermore, the CPU times required for CDVR calculations with or
without the use of quadrature optimized SPFs are roughly equal.
For some basis sets, the calculations with quadrature optimized
SPFs are faster than ones without. Thus, the effort for the computation
of quadrature optimized SPFs is apparently irrelevant. Variations in
integration steps sizes caused by the slight differences in the
propagated wavefunction are more relevant for the numerical effort. 

For comparison, CPU times required by the orginal non-hierarchical
CVDR approach are also shown in Fig.7. These CPU times increase more
rapidly with the size of the SPF basis consistent with the
\mbox{(\it total number of A coefficients\rm)$^{6/3}$} scaling law
expected and exceed those required by the revised CDVR
approach already for the smallest basis considered. For the moderately
sized B2 basis with about 70,000 A-coefficients, e.g., the difference
in the computation times amounts to about factor of five.

\section{Conclusions and outlook \label{sec7}}

The revised version of the non-hierarchical CDVR approach introduced
in this work enables highly efficient multi-layer MCTDH calculations
using general PESs of arbitrary structure. Compared to the
original non-hierarchical CDVR approach, the revised scheme reduces
the numerical effort while retaining the accuracy of quadrature.
Calculations with the revised CDVR exhibit the same computational scaling
with the size of the SPF basis sets as multi-layer MCTDH calculations employing
Hamiltonians in SOP form: for an ideally structure tree
with three edges at each node and $n$ SPF at each edge, the required
CPU time scales as $n^4$. Furthermore, the revised
CDVR approach eliminates artifacts resulting from the appearance  
of projections on the space spanned by the single-hole function
present in the original version of the non-hierarchical CDVR.

Calculations studying the non-adiabatic dynamics of photoexcited
pyrazine in full dimensionality (24D) impressively demonstrated
the numerical efficiency of the revised non-hierarchical CDVR approach.
Although this system is particularly well suited for multi-layer MCTDH calculations
exploiting the SOP structure of the diabatic potential, 
CDVR and SOP calculations require comparable CPU times. This result
highlights the potential of the revised CDVR scheme for multi-layer MCTDH calculations
on complex ab initio PESs, where a SOP representation is not readily available.

Furthermore, a revised scheme for redefining unoccupied SPFs is introduced
that systematically increases the accuracy of the CDVR quadrature without
incurring significant additional computational cost. In principle, this
approach enables systematic convergence of the quadrature to arbitrary accuracy.

Considering the numerical efficiency achieved by the revised
non-hierarchical CDVR, interesting new lines of development could been
envisioned. Employing on the fly-based techniques to obtain the
potential energies on the sparse grids required by the
non-hierarchical CDVR could be one direction.
Growing SOP potentials based on the relevant potential energy
values obtained by successive MCTDH calculations combining CDVR and SOP
representations of the potential are another option.

\section*{Acknowledgments}
The author wants to thank Hannes Hoppe for many stimulating discussions.

\section*{Data availability}
The data that supports the findings of this study is available within the 
article.\\

\section*{Appendix A: Derivation of the optimal 
$\Delta \hat{V}_{t,rev}^{\lambda|\lambda \circ \kappa}$ }

In the revised CDVR scheme, the action of the operator
$\Delta \hat{V}_{t}^{\lambda|\lambda \circ \kappa} \hat{H}_t$
on the system's wavefunction is approximated by the
action of the revised operator
$\Delta \hat{V}_{t,rev}^{\lambda|\lambda \circ \kappa} \hat{H}_t$.
The optimal choice for
$\Delta \hat{V}_{t,rev}^{\lambda|\lambda \circ \kappa}$ can be
obtained by 
\begin{align}
\delta \left\| \left( \Delta \hat{V}_{t}^{\lambda|\lambda \circ \kappa}
- \Delta \hat{V}_{t,rev}^{\lambda|\lambda \circ \kappa} \right)
\hat{H}_t \Psi \right\|^2=0~.
\end{align}
For hermitian
$\Delta \hat{V}_{t}^{\lambda|\lambda \circ \kappa}$
and $\Delta \hat{V}_{t,rev}^{\lambda|\lambda \circ \kappa}$, the
variation with respect to
$\Delta \hat{V}_{t,rev}^{\lambda|\lambda \circ \kappa}$ yields
\begin{align}
0=\delta \langle \hat{H}_t \Psi | &
-\Delta \hat{V}_{t}^{\lambda|\lambda \circ \kappa}
\delta \Delta \hat{V}_{t,ref}^{\lambda|\lambda \circ \kappa}
\nonumber\\&
-\delta \Delta \hat{V}_{t,ref}^{\lambda|\lambda \circ \kappa}
\Delta \hat{V}_{t}^{\lambda|\lambda \circ \kappa}
\nonumber\\&
+\Delta \hat{V}_{t,ref}^{\lambda|\lambda \circ \kappa}
\delta \Delta \hat{V}_{t,ref}^{\lambda|\lambda \circ \kappa}
\nonumber\\&
+\delta \Delta \hat{V}_{t,ref}^{\lambda|\lambda \circ \kappa}
\Delta \hat{V}_{t,ref}^{\lambda|\lambda \circ \kappa}
| \hat{H}_t \Psi \rangle 
~.
\label{eqA1}
\end{align}
Varying
\begin{align}
\Delta \hat{V}_{t,rev}^{\lambda|\lambda \circ \kappa}= \sum_{i,j}
| \xi^{\lambda |\lambda \circ \kappa}_{i} \rangle ~
\Delta \hat{V}_{t;ij}^{\lambda|\lambda \circ \kappa} ~
\langle \xi^{\lambda |\lambda \circ \kappa}_{j}|
\label{eqA2}
\end{align}
with respect to the $\Delta \hat{V}_{t;ij}^{\lambda|\lambda \circ \kappa}$,
the optimal $\Delta \hat{V}_{t;ij}^{\lambda|\lambda \circ \kappa}$
are thus given by 
\begin{align}
\langle \hat{H}_t \Psi | &
\Delta \hat{V}_{t}^{\lambda|\lambda \circ \kappa}
| \xi^{\lambda |\lambda \circ \kappa}_{i} \rangle 
\langle \xi^{\lambda |\lambda \circ \kappa}_{j}|
\nonumber\\& +
| \xi^{\lambda |\lambda \circ \kappa}_{i} \rangle 
\langle \xi^{\lambda |\lambda \circ \kappa}_{j}|
\Delta \hat{V}_{t}^{\lambda|\lambda \circ \kappa}
\nonumber\\& -
\Delta \hat{V}_{t,rev}^{\lambda|\lambda \circ \kappa}
| \xi^{\lambda |\lambda \circ \kappa}_{i} \rangle 
\langle \xi^{\lambda |\lambda \circ \kappa}_{j}|
\nonumber\\& -
| \xi^{\lambda |\lambda \circ \kappa}_{i} \rangle 
\langle \xi^{\lambda |\lambda \circ \kappa}_{j}|
{\Delta \hat{V}_{t,rev}^{\lambda|\lambda \circ \kappa}}
| \hat{H}_t \Psi \rangle =0~.
\label{eqA3}
\end{align}
Inserting the definitions of
$\Delta \hat{V}_{t}^{\lambda|\lambda \circ \kappa}$
(see Eq.(\ref{eq31}))
and $\Delta \hat{V}_{t,rev}^{\lambda|\lambda \circ \kappa}$,
the above equation can be rewritten as
\begin{align}
\langle \hat{H}_t \Psi | &
\sum_{n,m,l} | \Xi^{\lambda |\lambda \circ \kappa}_{nm} \rangle 
\Delta V_{t;nmil}^{\lambda|\lambda \circ \kappa}
\langle \Xi^{\lambda |\lambda \circ \kappa}_{jl}|
\nonumber\\& +
\sum_{n,m,l} | \Xi^{\lambda |\lambda \circ \kappa}_{il} \rangle 
\Delta V_{t;jlnm}^{\lambda|\lambda \circ \kappa}
\langle \Xi^{\lambda |\lambda \circ \kappa}_{nm}|
\nonumber\\& -
\sum_n | \xi^{\lambda |\lambda \circ \kappa}_{n} \rangle 
\Delta V_{t,ni}^{\lambda|\lambda \circ \kappa}
\langle \xi^{\lambda |\lambda \circ \kappa}_{j}|
\nonumber\\& -
\sum_n | \xi^{\lambda |\lambda \circ \kappa}_{i} \rangle 
\Delta V_{t,jn}^{\lambda|\lambda \circ \kappa}
\langle \xi^{\lambda |\lambda \circ \kappa}_{n}
| \hat{H}_t \Psi \rangle =0 ~.
\label{eqA3}
\end{align}
With the definitions introduced in Eqs.(\ref{eq36})
and (\ref{eq37}), one finds
\begin{align}
0=& ~ {X_{t;ji}^{\lambda|\lambda \circ \kappa}}^*+
X_{t;ij}^{\lambda|\lambda \circ \kappa}
\nonumber\\ &-
\sum_{n,m} u_{t;nm}^{\lambda|\lambda \circ \kappa}
~ r_{t;m}^{\lambda|\lambda \circ \kappa}
~ {u_{t;jm}^{\lambda|\lambda \circ \kappa}}^*
\cdot \Delta V_{t,ni}^{\lambda|\lambda \circ \kappa}
\nonumber\\ &-
\sum_{n,m} u_{t;im}^{\lambda|\lambda \circ \kappa}
~ r_{t;m}^{\lambda|\lambda \circ \kappa}
~ {u_{t;nm}^{\lambda|\lambda \circ \kappa}}^*
\cdot \Delta V_{t,jn}^{\lambda|\lambda \circ \kappa}~.
\label{eqA4}
\end{align}
Since the $\mathbf{u}_{t}^{\lambda|\lambda \circ \kappa}$ are
unitary matrices, multiplication with
${u_{t;ia}^{\lambda|\lambda \circ \kappa}}^*
u_{t;jb}^{\lambda|\lambda \circ \kappa}$ and summation over $i$ and
$j$ yields  
\begin{align}
0=&\sum_{i,j} {u_{t;ia}^{\lambda|\lambda \circ \kappa}}^*                                                                        
u_{t;jb}^{\lambda|\lambda \circ \kappa} \cdot
\left( {X_{t;ji}^{\lambda|\lambda \circ \kappa}}^*+
X_{t;ij}^{\lambda|\lambda \circ \kappa} \right)
\nonumber\\ &-
\sum_{n,i} u_{t;nb}^{\lambda|\lambda \circ \kappa}
~ r_{t;b}^{\lambda|\lambda \circ \kappa}
~ {u_{t;ia}^{\lambda|\lambda \circ \kappa}}^*
\cdot \Delta V_{t,ni}^{\lambda|\lambda \circ \kappa}
\nonumber\\ &-
\sum_{n,j} u_{t;jb}^{\lambda|\lambda \circ \kappa}
~ r_{t;a}^{\lambda|\lambda \circ \kappa}
~ {u_{t;na}^{\lambda|\lambda \circ \kappa}}^*
\cdot \Delta V_{t,jn}^{\lambda|\lambda \circ \kappa}~.
\label{eqA5}
\end{align}
Relabeling the summation indices, the two last terms
can be combined:
\begin{align}
0=&\sum_{k,l} {u_{t;ka}^{\lambda|\lambda \circ \kappa}}^*                                                                        
u_{t;lb}^{\lambda|\lambda \circ \kappa} \cdot
\left( {X_{t;lk}^{\lambda|\lambda \circ \kappa}}^*+
X_{t;kl}^{\lambda|\lambda \circ \kappa} \right)
\nonumber\\ &-
(r_{t;a}^{\lambda|\lambda \circ \kappa}
+r_{t;b}^{\lambda|\lambda \circ \kappa}) \sum_{n,m} 
u_{t;nb}^{\lambda|\lambda \circ \kappa}
~ {u_{t;ma}^{\lambda|\lambda \circ \kappa}}^*
~ \Delta V_{t,nm}^{\lambda|\lambda \circ \kappa}~.
\label{eqA5}
\end{align}
The final result given in the main part of the article,
Eq.(\ref{eq35}), is then obtain by multiplication with
\begin{align}
\frac{{u_{t;ib}^{\lambda|\lambda \circ \kappa}}^*
  ~ u_{t;ja}^{\lambda|\lambda \circ \kappa}}
{r_{t;a}^{\lambda|\lambda \circ \kappa}
+r_{t;b}^{\lambda|\lambda \circ \kappa}}
\end{align}
and summation over $a$ and $b$.

\section*{Appendix B: Separable potentials}

For separable potentials, the quadrature provided by the revised CDVR
is accurate even if the SPF basis set sizes are not converged. To prove
this feature, it is sufficient to prove that the revised CDVR is accurate
for potentials that dependent only on a single coordinate: since the
CDVR potential operator is a linear function of the potential, the
revised CDVR describes a sum of potentials correctly if each term in the
sum is described correctly.

Now we consider a potential $V_t$ that depends only a single coordinate
present in the coordinate set $q^{\lambda}$ associated with the
node $(\lambda)$ and investigate the potential corrections
$\Delta \hat{V}_{t,rev}^{\lambda \circ \kappa|\lambda}$.
Since the potential depends only on $q^\lambda$, $\hat{V}^{\lambda}_{t,\kappa}$
does not act on $|\xi_{l}^{\lambda|\lambda \circ \kappa} \rangle$
and $V_t({\boldsymbol \Xi}^{\lambda \circ \kappa | \lambda}_{k,i})
=V_t({\boldsymbol \xi}^{\lambda \circ \kappa | \lambda}_{k})$.
Then the $X_{t;lk}^{\lambda \circ \kappa|\lambda}$ in Eq.\ref{eq40} can be
simplified:
\begin{align}
X_{t;lk}^{\lambda \circ \kappa|\lambda}=&  
- \langle \Psi | \xi_{l}^{\lambda \circ \kappa|\lambda} \rangle
V_t({\boldsymbol \xi}^{\lambda \circ \kappa | \lambda}_{k})
\langle \xi_{k}^{\lambda \circ \kappa|\lambda} | \Psi \rangle 
\nonumber\\ &
+ \langle \Psi | \xi_{l}^{\lambda \circ \kappa|\lambda} \rangle
\sum_{n}
\sum_I \langle \xi^{\lambda \circ \kappa|\lambda}_{k}|\Xi^{\lambda}_I \rangle
\nonumber\\ & \quad
\cdot \langle \Xi^{\lambda}_I | \hat{V}^{\lambda}_{t,\kappa}|
\xi^{\lambda \circ \kappa|\lambda}_{n} \rangle
\langle \xi_{n}^{\lambda \circ \kappa|\lambda} | \Psi \rangle
\nonumber\\ =&  
\sum_n \langle \Psi | \xi_{l}^{\lambda \circ \kappa|\lambda} \rangle
\langle \xi_{n}^{\lambda \circ \kappa|\lambda} | \Psi \rangle
\nonumber\\ &\cdot
\langle \xi^{\lambda \circ \kappa|\lambda}_{k}|
- V_t({\boldsymbol \xi}^{\lambda \circ \kappa | \lambda}_{k}) +
\hat{V}^{\lambda}_{t,\kappa}| \xi_{n}^{\lambda \circ \kappa|\lambda}  \rangle
\nonumber\\ =&  
\sum_{n,m}
u_{t;lm}^{\lambda \circ \kappa|\lambda} \cdot
r_{t;m}^{\lambda \circ \kappa|\lambda}
\cdot {u_{t;nm}^{\lambda \circ \kappa|\lambda}}^*
\nonumber\\ &\cdot
\langle \xi^{\lambda \circ \kappa|\lambda}_{k}|
- V_t({\boldsymbol \xi}^{\lambda \circ \kappa | \lambda}_{k}) +
\hat{V}^{\lambda}_{t,\kappa}| \xi_{n}^{\lambda \circ \kappa|\lambda}  \rangle ~.
\end{align}
Analogously, one finds
\begin{align}
{X_{t;kl}^{\lambda \circ \kappa|\lambda}}^*=&  
\sum_{n,m}
u_{t;nm}^{\lambda \circ \kappa|\lambda} \cdot
r_{t;m}^{\lambda \circ \kappa|\lambda}
\cdot {u_{t;km}^{\lambda \circ \kappa|\lambda}}^*
\nonumber\\ &\cdot
\langle \xi^{\lambda \circ \kappa|\lambda}_{n}|
- V_t({\boldsymbol \xi}^{\lambda \circ \kappa | \lambda}_{l}) +
\hat{V}^{\lambda}_{t,\kappa}| \xi_{l}^{\lambda \circ \kappa|\lambda}  \rangle ~.
\end{align}
These two results can be combined to obtain
\begin{align}
C_{t;ij}^{\lambda \circ \kappa|\lambda}=&
\sum_{l,k} {u_{t;li}^{\lambda \circ \kappa|\lambda}}^* \cdot
u_{t;kj}^{\lambda \circ \kappa|\lambda} \cdot
({X_{t;kl}^{\lambda \circ \kappa|\lambda}}^* +
X_{t;lk}^{\lambda \circ \kappa|\lambda})
\nonumber\\ =&
\sum_{n,l}
u_{t;nj}^{\lambda \circ \kappa|\lambda} \cdot
r_{t;j}^{\lambda \circ \kappa|\lambda}
\cdot {u_{t;li}^{\lambda \circ \kappa|\lambda}}^*
\nonumber\\ &\cdot
\langle \xi^{\lambda \circ \kappa|\lambda}_{n}|
- V_t({\boldsymbol \xi}^{\lambda \circ \kappa | \lambda}_{l}) +
\hat{V}^{\lambda}_{t,\kappa}| \xi_{l}^{\lambda \circ \kappa|\lambda}  \rangle
\nonumber\\ &+
\sum_{n,k}
u_{t;kj}^{\lambda \circ \kappa|\lambda} \cdot
r_{t;i}^{\lambda \circ \kappa|\lambda}
\cdot {u_{t;ni}^{\lambda \circ \kappa|\lambda}}^*
\nonumber\\ &\cdot
\langle \xi^{\lambda \circ \kappa|\lambda}_{k}|
- V_t({\boldsymbol \xi}^{\lambda \circ \kappa | \lambda}_{k}) +
\hat{V}^{\lambda}_{t,\kappa}| \xi_{n}^{\lambda \circ \kappa|\lambda}  \rangle
\nonumber\\ =&
\sum_{n,m}
u_{t;mj}^{\lambda \circ \kappa|\lambda} \cdot
(r_{t;i}^{\lambda \circ \kappa|\lambda}+r_{t;j}^{\lambda \circ \kappa|\lambda})
\cdot {u_{t;ni}^{\lambda \circ \kappa|\lambda}}^*
\nonumber\\ &\cdot
\langle \xi^{\lambda \circ \kappa|\lambda}_{m}|
- V_t({\boldsymbol \xi}^{\lambda \circ \kappa | \lambda}_{m}) +
\hat{V}^{\lambda}_{t,\kappa}| \xi_{n}^{\lambda \circ \kappa|\lambda}  \rangle
\end{align}
Inserting this result into Eq.\ref{eq35}, the final result for the
potential correction matrix reads
\begin{align}
\Delta V_{t;ij}^{\lambda \circ \kappa|\lambda}=&
\langle \xi^{\lambda \circ \kappa|\lambda}_{i}|
- V_t({\boldsymbol \xi}^{\lambda \circ \kappa | \lambda}_{i}) +
\hat{V}^{\lambda}_{t,\kappa}| \xi_{j}^{\lambda \circ \kappa|\lambda}  \rangle~.
\end{align}
Consequently, the operator associated with the potential correction
is the revised CDVR for a potential depending only on $q^\lambda$,  
\begin{align}
\Delta \hat{V}_{t,rev}^{\lambda \circ \kappa|\lambda}=&
- \hat{V}_t^{\lambda \circ \kappa | \lambda} + \hat{V}^{\lambda}_{t,\kappa}~,
\end{align}
correctly compensates for all errors resulting from the use of the
scarce grid ${\boldsymbol \Xi}^{\lambda \circ \kappa}_N$ at node $(\lambda \circ \kappa)$.

\end{document}